\RequirePackage{rotating}

\documentclass[a4paper,fleqn,usenatbib]{mnras}

\hypersetup{colorlinks=true,linkcolor=blue,citecolor=blue,filecolor=blue,urlcolor=blue}
\usepackage{newtxtext,newtxmath}

\usepackage{graphicx}

\newcommand {\be} {\begin{equation}}
\newcommand {\ee} {\end{equation}}

\usepackage[normalem]{ulem}

\usepackage{color}
\usepackage{upgreek}	
\usepackage{threeparttablex}
\usepackage{pdflscape}  
\usepackage{afterpage}
\usepackage{capt-of}
\usepackage{lipsum}
\usepackage{subcaption}
\usepackage{rotating}
\usepackage[T1]{fontenc}
\usepackage{ae,aecompl}

\usepackage{graphicx}	
\usepackage{amsmath}	
\newcommand{\angstrom}{\textup{\AA}}
\usepackage{amssymb}	
\usepackage{minibox}

\title[The SEDs of very massive RL and 
RQ AGN]{Comparison of SEDs of very massive radio-loud and radio-quiet AGN}

\author[Gupta et al.]{
Maitrayee Gupta,$^{1}$\thanks{E-mail: mgupta@camk.edu.pl}
Marek Sikora,$^{1}$
Katarzyna Rusinek$^{1}$
\\
$^{1}$Nicolaus Copernicus Astronomical Center, Bartycka 18, 00-716 Warsaw, Poland}

\date{Accepted XXX. Received YYY; in original form ZZZ}

\pubyear{2019}

\begin{document}
\label{firstpage}
\pagerange{\pageref{firstpage}--\pageref{lastpage}}
\maketitle

\begin{abstract}

The main objective of this work is to establish and interpret the dominant spectral components and their differences in radio-loud (RL) and radio-quiet (RQ) AGN with very massive black holes, and accreting at moderate rates. Such a sample is selected from the \textit{Swift}/BAT catalogue of AGN having determined optical spectra types and hosting black holes with masses $\rm > 10^{8.5}\,M_{\odot}$. We confirm our previous results, that radio loudness distribution of \textit{Swift}/BAT AGN is bimodal and that radio galaxies are about two times X-ray louder than their radio-quiet counterparts. We show that the average X-ray loudness (defined as a ratio of luminosity in the $14-195$ keV band to that at 12 $\upmu$m) of Type 1 and Type 2 AGN is very similar. This similarity holds for both RL and RQ subsamples and indicates negligible dependence of the observed X-ray luminosities on the inclination angle in both populations.

In both the radiative output is dominated by mid-IR and hard X-ray components, and relatively weak UV luminosities indicate large amounts of dust in polar regions.

\end{abstract}



\section{INTRODUCTION}
\label{sec:intro}

One of the biggest puzzles of the AGN phenomenon is its large
 diversity regarding the jet production efficiency. This is particularly apparent in AGN
with higher accretion rates, where power of a jet traced by radio luminosity can be compared with accretion power traced by optical  
luminosity of accretion discs or IR luminosity resulting from reprocessing of optical-UV radiation by dusty obscurers.  At each Eddington ratio the radio loudness spans 
by 3 to 4 orders of magnitude \citep[e.g.][]{2007ApJ...658..815S}. In the case of 
Blandford-Znajek mechanism of  a jet production such a diversity is expected to be related  to the range of values of BH spins and magnetic fluxes threading them \citep{1977MNRAS.179..433B}.
Studies of differences in spectra of RL and RQ objects may help to establish 
whether central accumulation of large net magnetic fluxes in RL AGN proceeds 
during the AGN radiatively efficient phase or prior to it \citep{2013ApJ...765...62S, 
2013ApJ...764L..24S}.

Comparisons of multi-band spectra of RL and RQ AGN were performed in the past
for quasars \citep[e.g.][]{1994ApJS...95....1E,2006ApJS..166..470R,2011ApJS..196....2S} and for the samples composed  from broad-line 
radio galaxies  and Seyferts \citep[e.g.][]{1998MNRAS.299..449W,2011ApJ...740...29K}. 
The main radiative differences noted 
by these studies concern the X-ray properties. In particular, the RL objects
were found to have on average larger X-ray luminosities and harder X-ray 
spectra than the RQ ones. However noting that the objects included in the samples of compared RQ and RL objects were selected separately, that radio galaxies have on average larger BH masses and lower Eddington ratios than RQ Seyferts \citep[e.g.][]{2004MNRAS.353L..45M,2017ApJ...846...42K}, and that the slopes of the X-ray spectra depend on the Eddington ratio \citep[e.g.][]{2009MNRAS.399..349G,2017MNRAS.470..800T,2018ApJ...859..152S},
the claimed differences might be at least partially affected by the selection 
methods. Trying to avoid  biases resulting from separate selection
of RL and RQ samples and from having RL and RQ samples with 
different average BH masses and Eddington ratios, we decided to carry such 
comparisons selecting AGN from the BAT AGN Spectroscopic Survey catalogue 
\citep[BASS;][]{2017ApJ...850...74K} with similar  ranges of  BH masses and Eddington 
ratios.  First results of these  studies were presented by \cite{2018MNRAS.480.2861G}.
We have confirmed there that RL AGN are X-ray louder than RQ AGN, but did not find statistically significant difference between their X-ray spectral slopes. 

In this paper we extended our comparison studies of RL and RQ AGN by covering also other than hard X-ray spectral bands and  taking 
advantage of our sample including both Type 1 and Type 2 AGN.
The latter allowed us to verify an isotropy
of these radiative features, which are not expected to be affected via absorption by dusty molecular tori. This in particular concerns  hard X-rays of 
Compton-thin AGN and narrow emission lines. Furthermore, having the RL and RQ 
samples divided into Type 1 and Type 2 subsamples allowed us to make sure 
that observed in Type 1 AGN UV radiation is dominated by AGN.
In order to have reasonable sizes of statistically comparable
RL and RQ Type 1 and Type 2 subsamples we had to 
modify our previous selection criteria. The main change is that presently we 
include in our sample also those AGN, which before did not have known 
estimations of BH masses. 
We calculated their masses using the relation between their BH masses and the NIR 
luminosities of their host galaxies \citep{2003ApJ...589L..21M}. And in order 
to avoid a large scatter of BH masses estimated using different methods, 
the known masses of other AGN were recalculated using also this method (see Appendix \ref{app:mbh}).

The work presented here is organized as follows: in Section \ref{sec:sample} we describe 
procedure of selecting our sample; in Section \ref{sec:X-ray_prop} we present the 
results of our X-ray analysis; the Section \ref{sec:sed_distribution} deals with multi-band studies; 
in Section \ref{sec:discussion} we discuss our results in the broader context of AGN phenomenon; and the main results are  summarized in Section \ref{sec:summary}.

Throughout the paper we assume a $\Lambda$CDM cosmology with 
$\rm H_0 = 70 \ {\rm km} \ {\rm s}^{-1} \ {\rm Mpc}^{-1}$, $\rm \Omega_m=0.3$, and 
$\rm \Omega_{\Lambda}=0.70$.


\section{THE SAMPLES}
\label{sec:sample}

\begin{figure}
\centering
\includegraphics[scale=0.36]{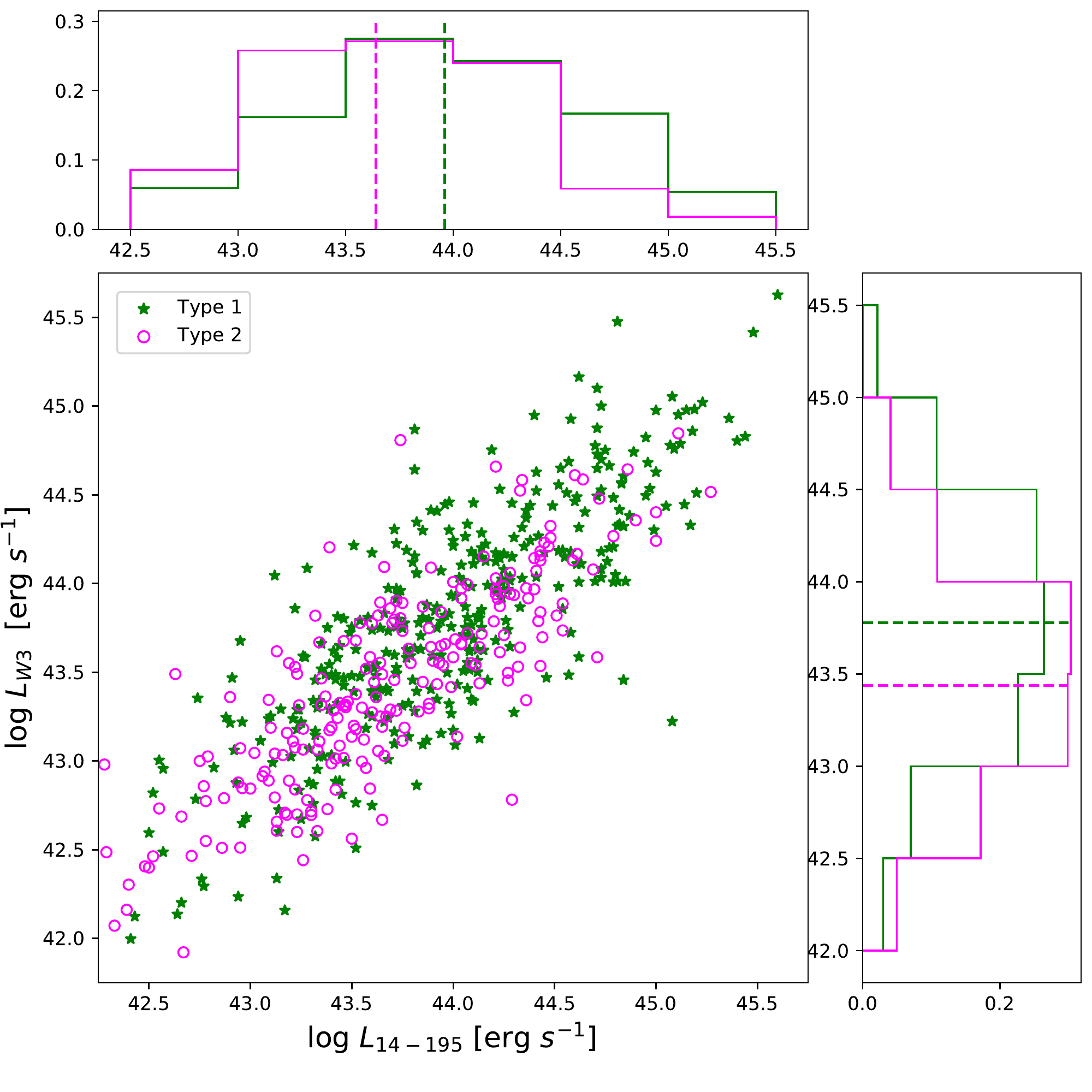}
\caption{The distribution of $\rm L_{\rm{14-195}}$ vs. $\rm L_{\rm{W3}}$ for the entire sample. Type 1 AGN which consists of Seyfert 1-1.9 galaxies are shown by green stars while the Type 2 AGN are marked by the pink circles. There is a significant overlap between the two samples. Also shown are the histograms of $\rm L_{\rm{14-195}}$ and $\rm L_{\rm{W3}}$ with their median values, presented as dashed lines. The median values of $\rm L_{\rm{14-195}}$ Type 1 and Type 2 are 43.96 and 43.64, respectively. The median values of $\rm L_{\rm{W3}}$ Type 1 and Type 2 are 43.78 and 43.44, respectively.}
\label{img_hist_w3_lx}
\end{figure}

\begin{figure*}
\centering
\includegraphics[width=\textwidth]{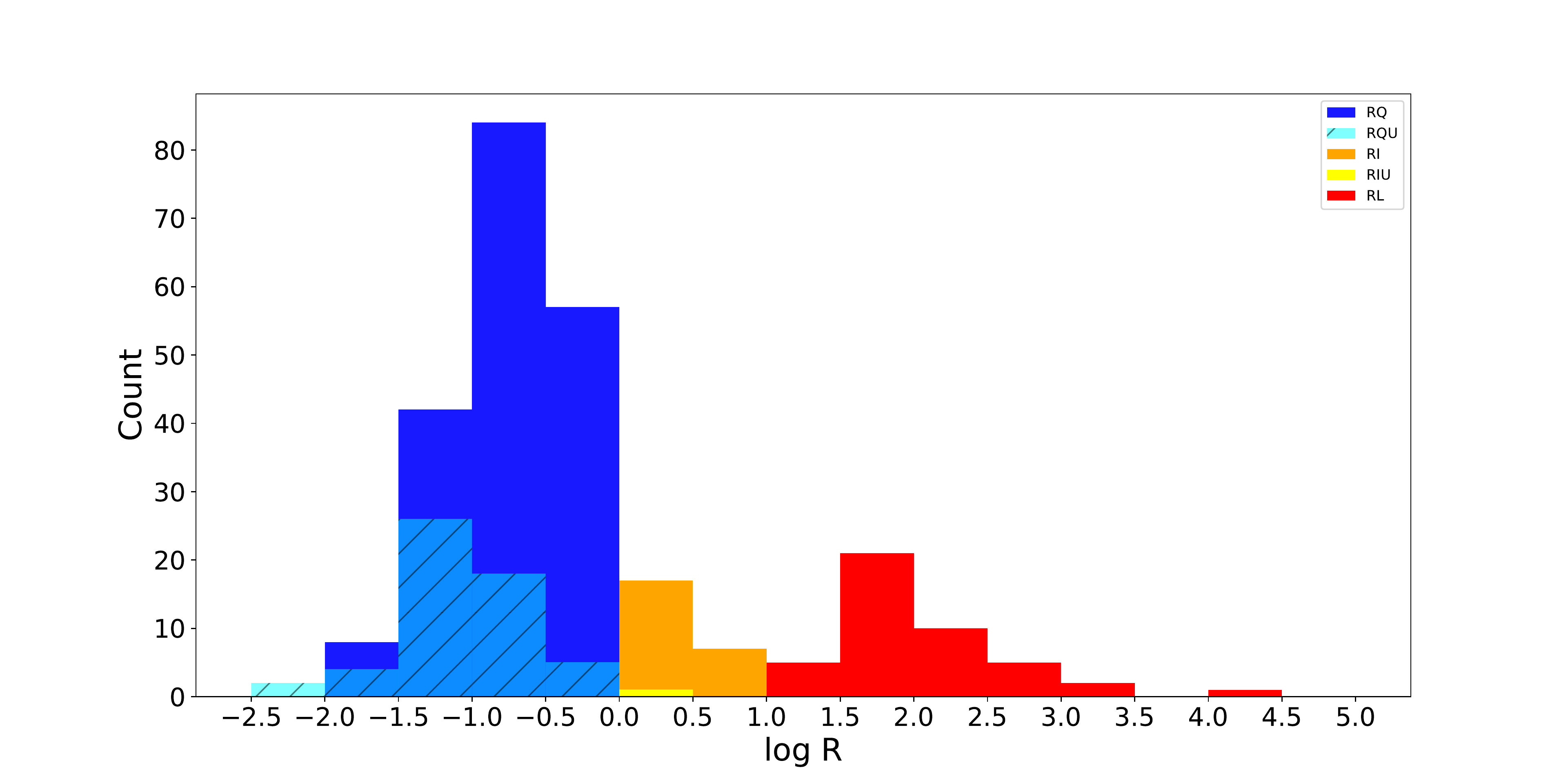}
\caption{The distribution of the radio loudness, $\rm R = F_{\rm 1.4}/F_{\rm \nu_{\rm W3}}$, after removing AGN with $\rm M_{\rm BH} < 10^{8.5}\,M_{\odot}$ is shown. Besides showing objects with radio detections and categorized as: $\rm R< 1$ as RQ in blue; $\rm 1 < R < 10$ as RI in yellow; and $\rm R > 10$ as RL in red, we also present the radio undetected sources classified as RQU in hatched blue and RIU in hatched yellow.}
\label{img_RL_histogram_mbh_nh_cutoff}
\end{figure*}

The primary sample of AGN selected in this work is the same as in \cite{2018MNRAS.480.2861G}. It is taken from the BAT AGN Spectroscopic Survey \citep{2017ApJS..233...17R} and then reduced by excluding blazars and  
Compton-thick (log $\rm N_{\rm H} > 24$) AGN. Such a sample, counting 664 objects,
was further reduced by excluding those AGN which do not have optical spectroscopic
classification in \cite{2017ApJ...850...74K}

Following this  we selected 394 Type 1 AGN and 232 Type 2 AGN, where the Type 1 sample includes all AGN having at least one broad Balmer emission line.
All AGN in these samples
have data on hard X-ray fluxes ($ \rm 14-195$ keV) from \textit{Swift}/Burst Alert Telescope 
\citep{2013ApJS..207...19B} and on mid-IR fluxes from Wide-field Infrared Survey
Explorer mission \citep[WISE;][]{2010AJ....140.1868W,2018MNRAS.480.2861G}.  
Distributions of their X-ray luminosities in the band $14-195$ keV and of 
mid-IR luminosities at $\rm \nu_{\rm W3} =  2.5 \times 10^{13}  Hz \equiv 12~\upmu m$, i.e.
$\rm L_{\rm W3} \equiv \nu_{\rm W3} L_{\rm \nu_{\rm W3}}$, are presented in Fig. \ref{img_hist_w3_lx}. They are 
calculated from fluxes using a standard cosmological formula, but ignoring the K-corrections since the redshifts for our AGN sample are small ($\rm z\leq0.35$) and the radiation spectral slopes in these bands are not very different from unity. As we can see in this figure, there is a significant overlap between the distributions of Type 1 and Type 2 AGN. It is confirmed by very low  Kolmogorov-Smirnov (KS) test statistic p-values of $2.26 \times 10^{-5}$ for the X-ray luminosities and $2 \times 10^{-7}$ for the MIR luminosity indicating that the two data samples come from the same distribution. This reasserts the unified scheme for AGN \citep{1995PASP..107..803U} and  affirms that hard X-rays and MIR are both produced quasi-isotropically. As a result we can treat these properties in the Type 1 and Type 2 AGN samples uniformly. 

The above Type 1 and Type 2 BAT-AGN subsamples have then been cross-matched
with NVSS \citep[National Radio Astronomy Observatory (NRAO) Very Large Array (VLA) Sky
Survey;][]{1998AJ....115.1693C} and SUMSS \citep[Sydney University Molonglo Sky
Survey;][]{1999AJ....117.1578B,2003MNRAS.342.1117M} radio catalogues using the same procedure as
in \cite{2018MNRAS.480.2861G}. Following this we divided our samples for 
radio-loud (RL), radio-intermediate (RI) and radio-quiet (RQ) AGN, where
the RL AGN are defined to have the radio loudness parameter $\rm R \equiv
L_{\rm 1.4}/L_{\rm \nu_{\rm {W3}}} > 10$ and RQ AGN are defined to have $\rm R < 1$.\footnote { Note that the radio loudness defined by us is about 10 times lower than the one introduced by \cite{1989AJ.....98.1195K} \citep[see section 2.4 in][]{2018MNRAS.480.2861G}. And the reason of using in our definition of radio loudness the MIR luminosity instead of optical luminosity is because at moderate and low accretion rates optical radiation can be significantly contaminated by the stellar radiation.}
  
We hereafter reject from our further analysis RI AGN and those radio-undetected objects which have radio loudness  upper limits $> 1$. This assures
of having two `clean' and well contrasted subsamples, those with $R>10$ where radio emission is 
expected to be strongly dominated by powerful jets, and those
with $R<1$, where the total radio flux can, additionally to weak jets \citep{2016ApJ...832..163S}, be contributed by star-forming regions \citep[e.g.][]{2011ApJ...739L..29K}, accretion disc winds
\citep[e.g.][]{2016MNRAS.459.3144Z}, and/or accretion disc coronae \citep[e.g.][]{2016MNRAS.459.2082R}.

Noting that in the samples of AGN with smaller BH masses, the RL AGN are poorly represented \citep{2004MNRAS.353L..45M,2017ApJ...846...42K,2018MNRAS.480.2861G}, we leave in our samples only AGN with BH masses larger than $\rm 10^{8.5}\,M_{\odot}$. Such BHs typically reside in giant ellipticals and their masses can be estimated by using galactic luminosities of old stellar populations \citep{2016ARA&A..54..597C} provided they dominate contribution to near-IR radiation \citep{2003ApJ...589L..21M,2007MNRAS.379..711G}. The latter condition is satisfied not only in Type 2 AGN, but also in  Type 1 AGN if accreting at not very high rates \citep[see e.g.][]{2006ApJS..166..470R}. As we demonstrate in Appendix \ref{app:mbh}, near-IR luminosities  of Type 1 AGN are only by a factor $1.47$ larger than near-IR luminosities of Type 2 AGN. Hence, after applying respective corrections, the $\rm M_{\rm BH} - L_{\rm NIR}$ relation can be applied also for Type 1 AGN in our sample.

The radio loudness histogram of our final sample (i.e. after removing AGN with $\rm M_{\rm BH} < 10^{8.5}\,M_{\odot}$) is shown in Fig. \ref{img_RL_histogram_mbh_nh_cutoff}. The sample consists of 290 AGN (315 AGN while including RI). Its RL subset has 44 objects (27 Type 1 and 17 Type 2) and the RQ subset has 246 objects (153 Type 1 and 93 Type 2).\footnote {The complete catalogue is available as supplementary material online.}

\section{Comparison of X-ray Properties}
\label{sec:X-ray_prop}

\subsection{X-ray loudness}
\label{subsec:X-ray_loud}

\begin{figure}
\centering
\includegraphics[scale=0.25]{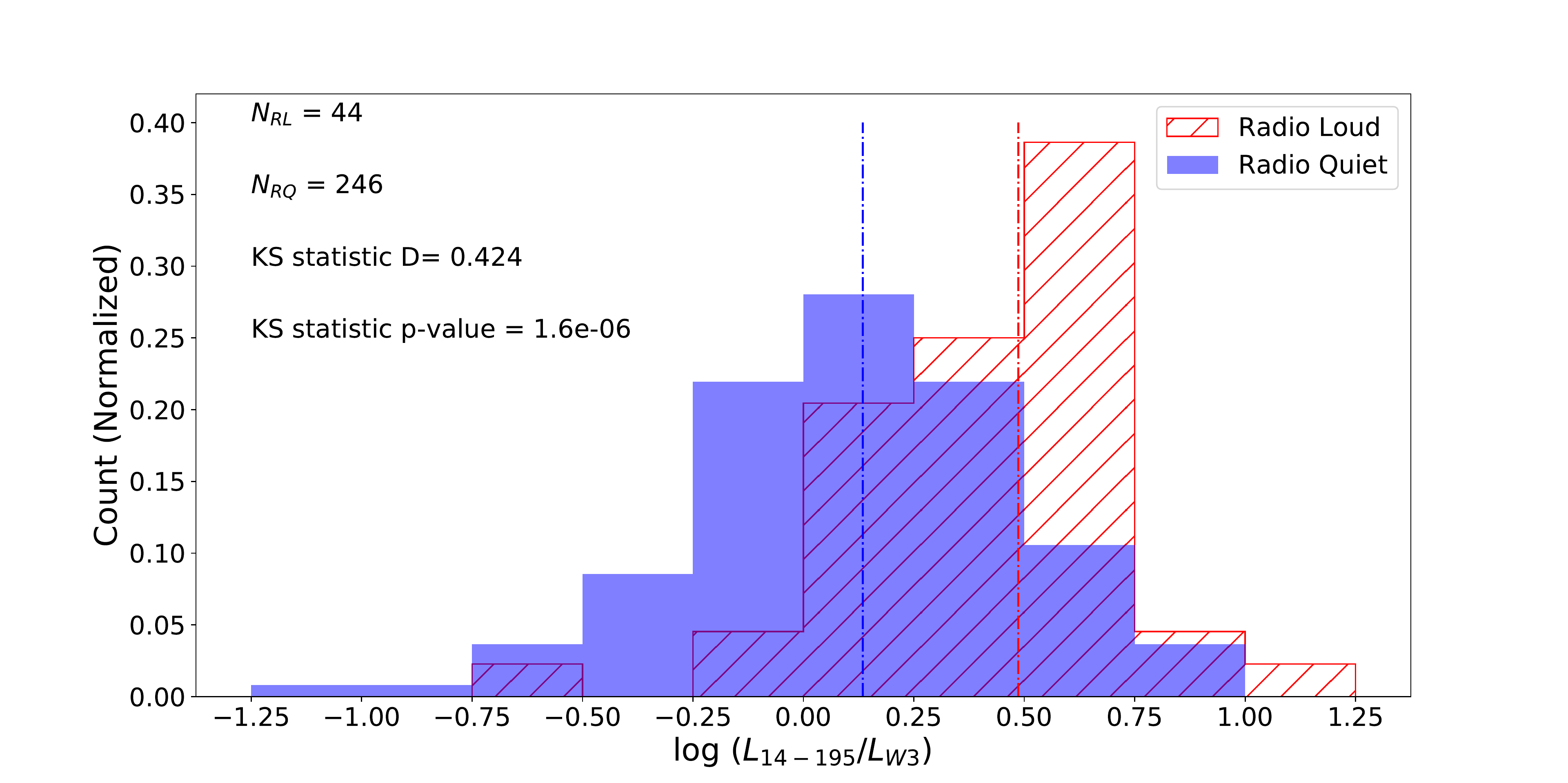}
\caption{The normalized distribution of the X-ray loudness $\rm L_{14-195}/L_{W3}$ for the RL (hatched red) and RQ (filled blue) samples. The median values of 0.49 and 0.13, respectively, are presented as dashed lines.}
\label{img_hist_hxlum}
\end{figure}

We define the X-ray loudness as the ratio of the hard X-ray luminosities in the band $14-195$ keV to the mid-IR luminosities at $\nu = \nu_{\rm W3}$. Distributions of X-ray loudness of RL and RQ AGN  are shown in Fig. \ref{img_hist_hxlum}.  As we can see these distributions have similar shapes, but are shifted against each other, with the median of X-ray loudness of RL AGN being about two times larger. This is in agreement with the result found in our previous work, despite having somewhat differently selected final sample.

\begin{figure*}
\begin{subfigure}{0.49\linewidth}
\centering
\includegraphics[scale=0.25]{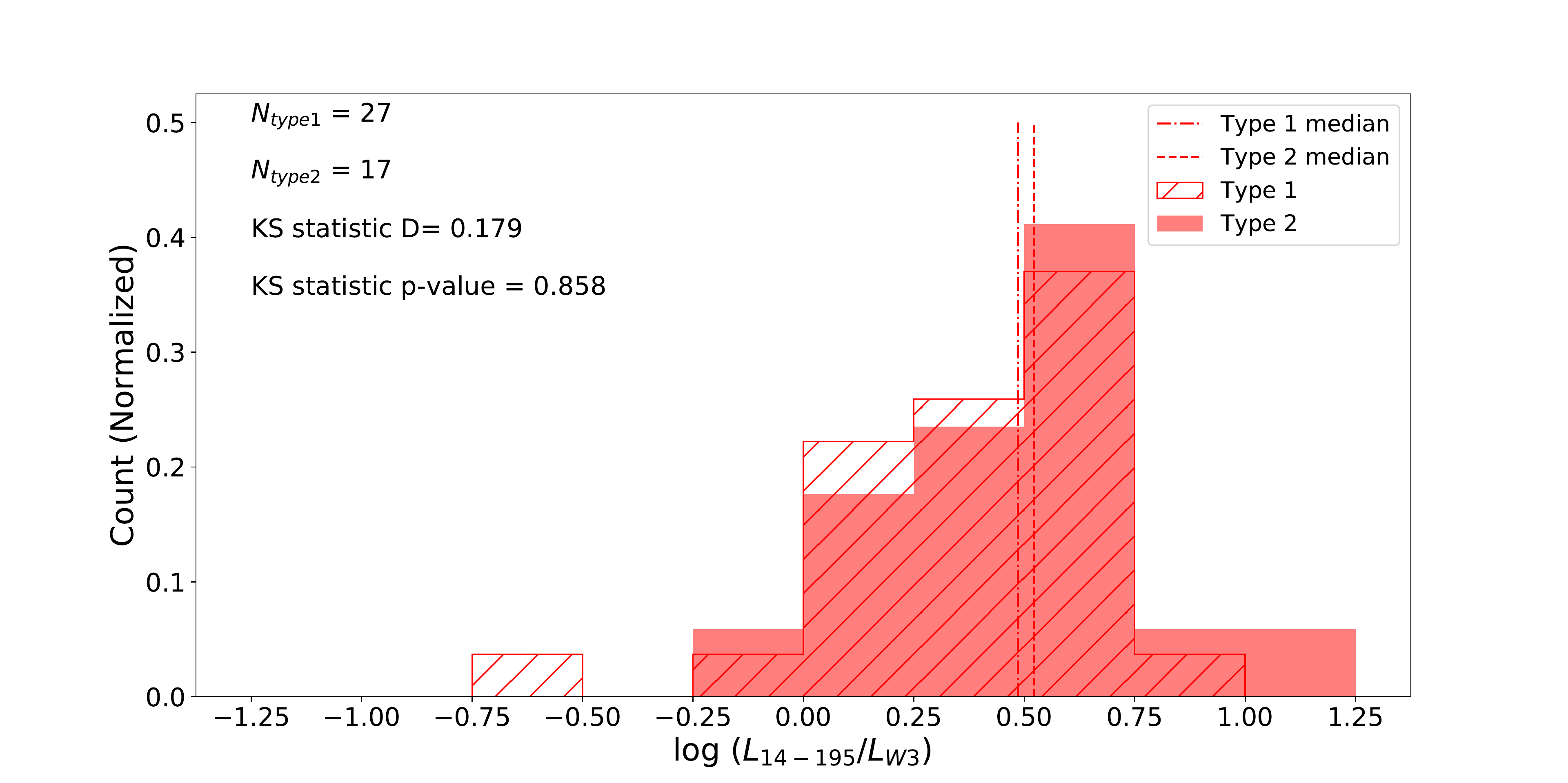}
\caption{}
\label{img_x_hard_hist_rl}
\end{subfigure}\hfill
\begin{subfigure}{0.49\linewidth}
\centering
\includegraphics[scale=0.25]{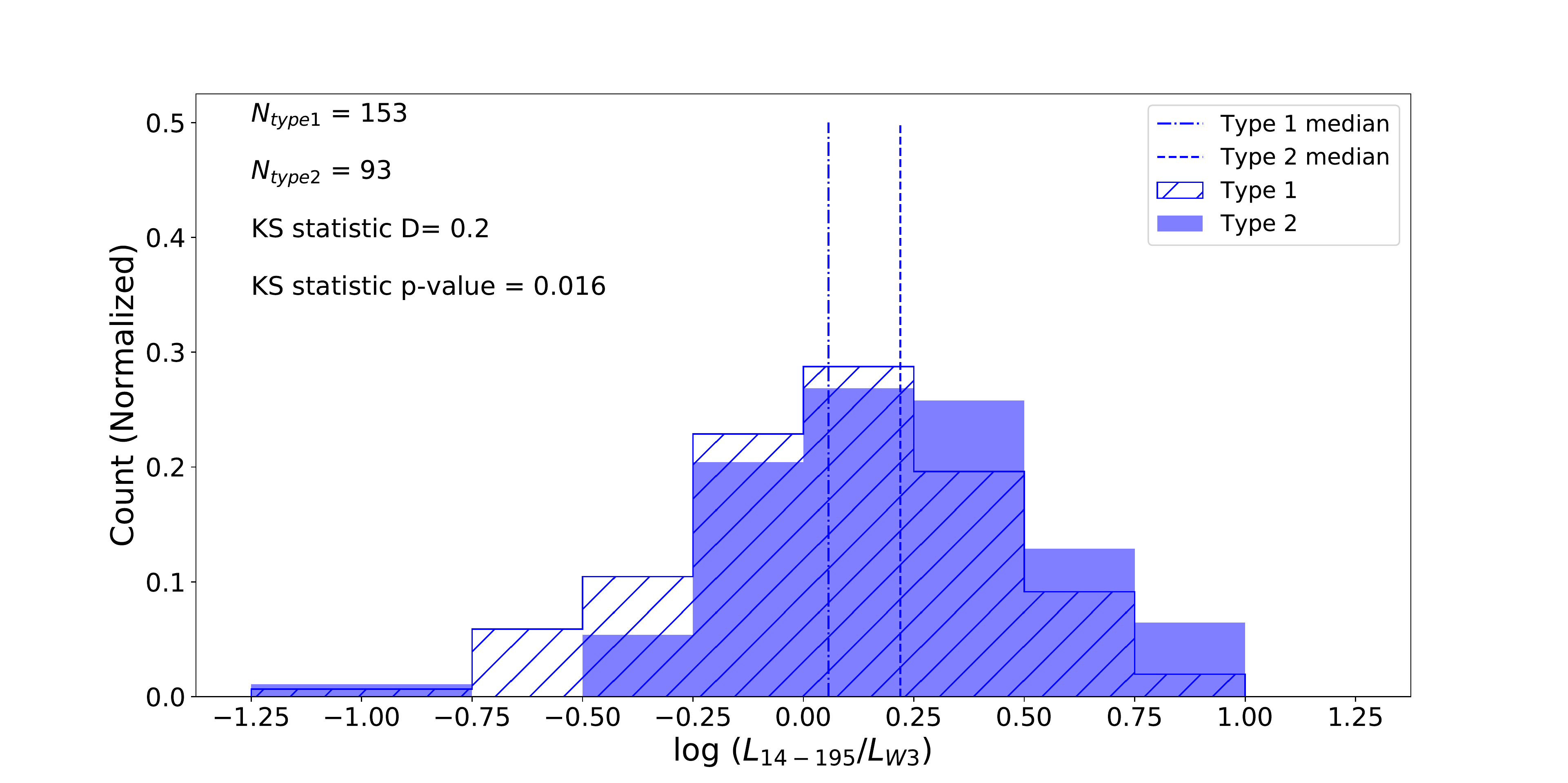}
\caption{}
\label{img_x_hard_hist_rq}
\end{subfigure}\hfill
\caption{The normalized distribution of the hard X-ray luminosity $\rm L_{14-195}$ for the RL (panel a) and RQ (panel b) samples. The luminosity has been normalized w.r.t. to the W3 luminosity. We present here the distributions of the Type 1 (hatched red/blue) and Type 2 (filled red/blue) subsamples of AGN. The median values of 0.49 and 0.52 for Type 1 and Type 2 RL AGN and 0.06 and 0.22 for Type 1 and Type 2 RQ AGN, respectively, are presented as dashed lines.}

\end{figure*}

\begin{figure*}
\begin{subfigure}{0.49\linewidth}
\centering
\includegraphics[scale=0.25]{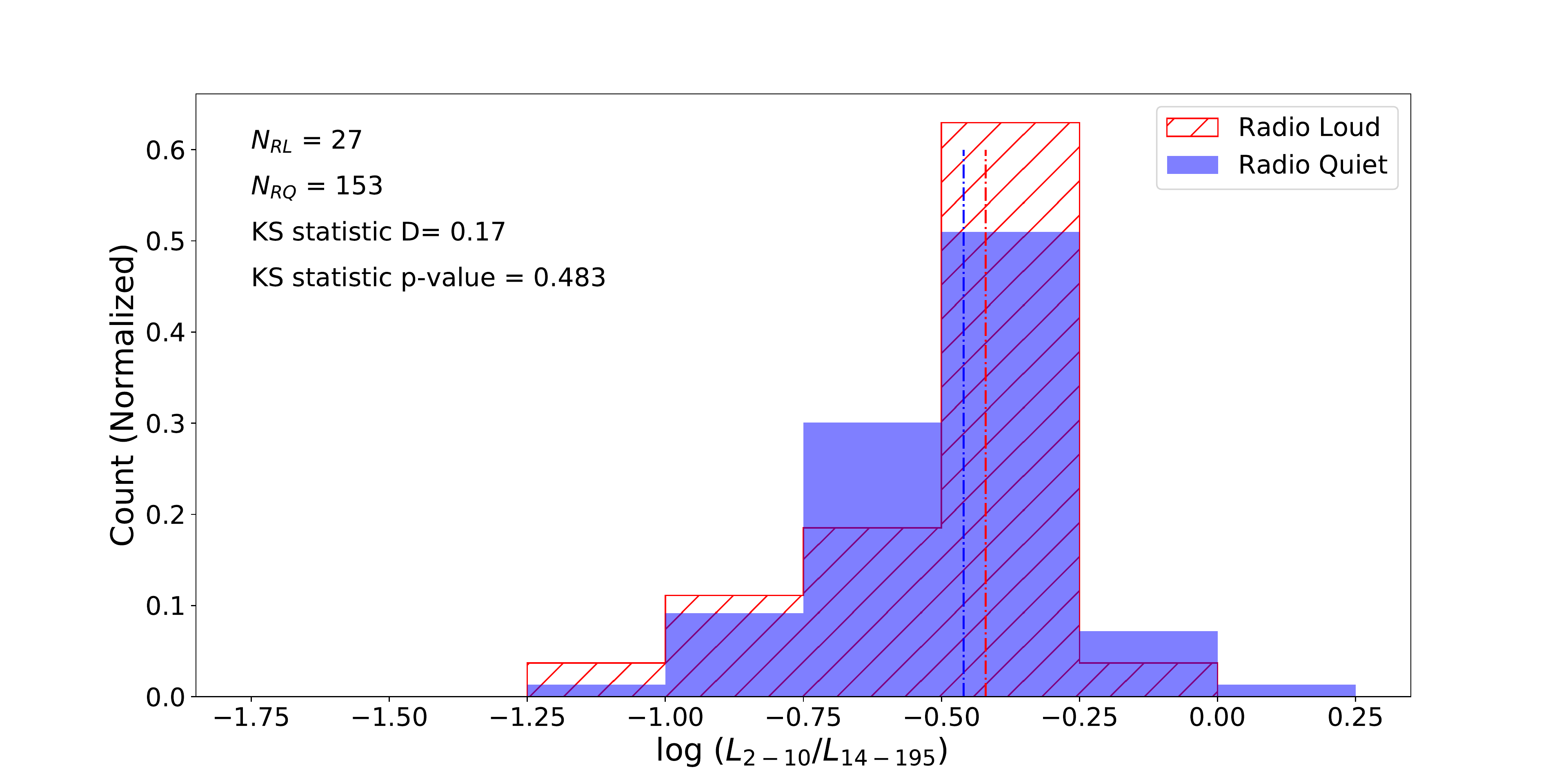}
\caption{}
\label{img_x_color_hist_ty1}
\end{subfigure}\hfill
\begin{subfigure}{0.49\linewidth}
\centering
\includegraphics[scale=0.25]{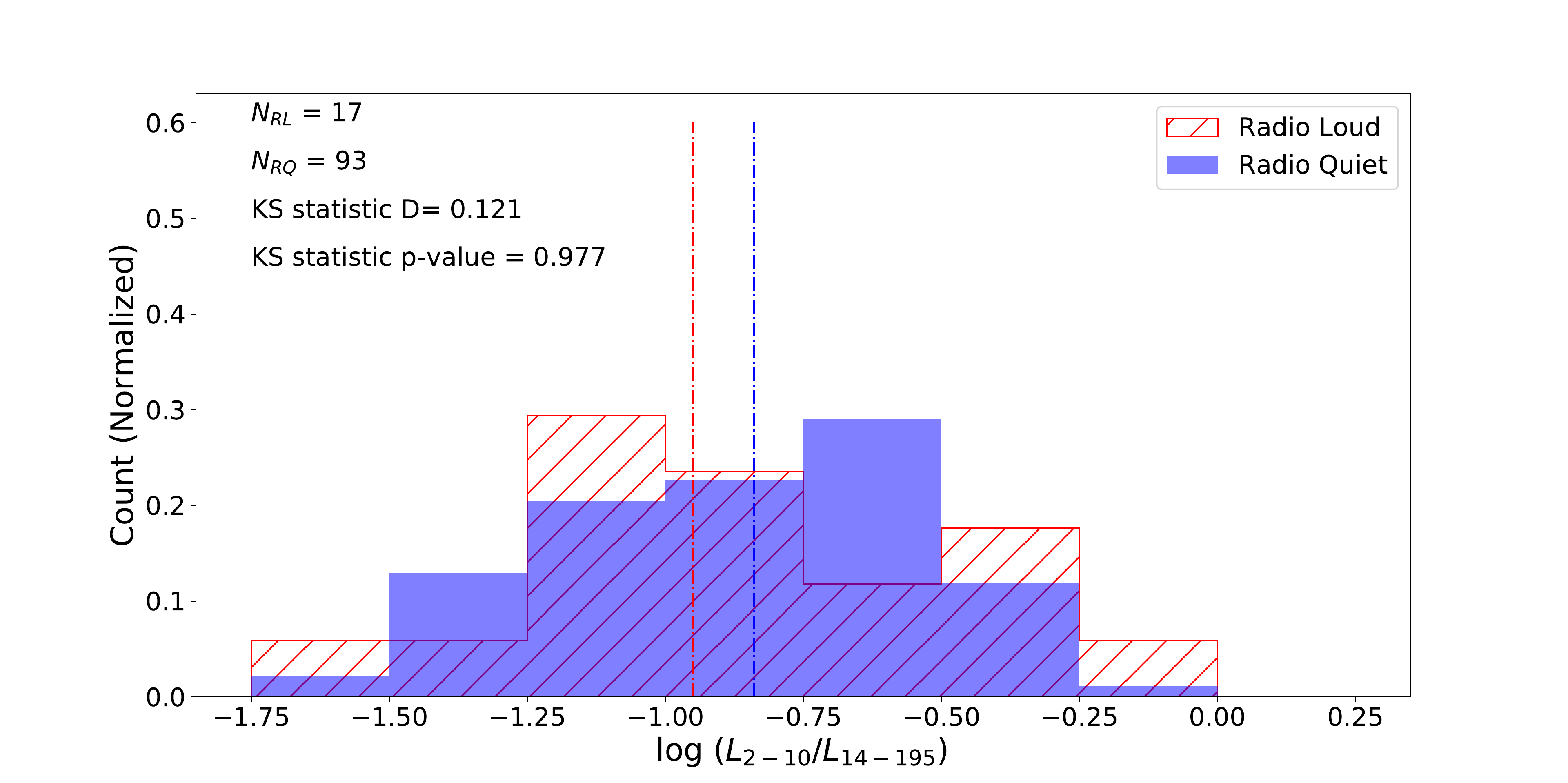}
\caption{}
\label{img_x_color_hist_ty2}
\end{subfigure}\hfill
\caption{The normalized distribution of the X-ray colour  $\rm L_{2-10}/L_{14-195}$ for the Type 1 (panel a) and Type 2 (panel b) samples. We present here the distributions of the RL (hatched red) and RQ (filled blue) subsamples of AGN. The median values of $-0.42$ and $-0.46$ for RL and RQ Type 1 AGN and $-0.95$ and $-0.84$ for RL and RQ Type 2 AGN, respectively, are presented as dashed lines.}
\end{figure*}

\subsection{Isotropy of hard X-rays}
\label{subsec:X-ray_isotropy}

As we argued in our previous paper \citep{2018MNRAS.480.2861G}, similar slopes
of high energy spectra and of their high energy breaks
in the radio-loud and radio-quiet \textit{Swift}/BAT AGN suggest same location
and mechanism of their production in both subsamples. Therefore
the larger X-ray loudness of RL AGN may not result from contribution
of jets to the hard X-ray spectra, but rather from larger efficiency 
of hard X-ray production in accretion flows in RL than in RQ AGN. Since the MIR radiation is expected to be isotropic \citep[see e.g.][]{2013ApJ...777...86L}, this premise can be verified by  checking, whether X-ray loudness depends on 
the angle of view, and if yes, whether it depends similarly in case of RL and 
RQ AGN. This can be done by noting that the Type 1 AGN are on average
observed at much smaller inclination angles than the Type 2 AGN.
Results of such a test are presented in the Fig. \ref{img_x_hard_hist_rl} and \ref{img_x_hard_hist_rq}. As we can see, the averaged X-ray loudness of Type 1 and Type 2 RL AGN are comparable with median values of 0.49 and 0.52 respectively. In addition their distributions are alike, a fact attested by the relatively large Kolmogorov-Smirnov (KS) test p-value of the null hypothesis being 0.858. We observe a similar trend for the RQ population with the distribution with the medians of the Type 1 and Type 2 subsets being at 0.06 and 0.22 respectively and K-S test p-value of 0.016. This strongly suggest similar regions of production of X-rays and similar underlying mechanisms in both Type 1 and Type 2 and for both the samples of RL and RQ AGN. These results strongly support our earlier premise that the X-ray emission 
in both RL and RQ AGN is dominated by the accretion flow and is 
quasi-isotropic.

\subsection{X-ray colours}
\label{subsec:X-ray_colour}

We define the X-ray colour as the ratio of X-ray luminosity in the $2-10$ keV band to the luminosity in the $14-195$ keV band -- $\rm L_{\rm 2-10}/L_{\rm 14-195}$. The data for the $2-10$ keV was obtained from \cite{2017ApJS..233...17R}. Fig. \ref{img_x_color_hist_ty1} and  \ref{img_x_color_hist_ty2} show the normalized distribution of the X-ray colour for the Type 1 (panel a) and Type 2 (panel b) samples. In each panel we have shown the distributions of the RL (hatched red) and RQ (filled blue) subsamples of AGN. The median values are also presented as dashed lines for each sample.
For the Type 1 sample we see very similar distributions for the RL and RQ subsets which is indicated by the high K-S test p value of 0.483. Their median values of $-0.42$ for the RL subset and $-0.46$ for the RQ subset are comparable.
We observe a similar phenomena in the Type 2 sample with RL and RQ subsample medians at $-0.95$ and $-0.84$ respectively and the K-S test p value of 0.977 indicating no statistically significant difference between RL and RQ AGN for either Type 1 or Type 2 populations.  
Note that harder X-ray colours in Type 2 RL and RQ AGN result from partial absorption of lower energy X-rays by molecular torus in Type 2 AGN.

\section{Spectral Energy Distribution}
\label{sec:sed_distribution}

We compare the SEDs of RL and RQ AGN using photometric data
covering  MIR, NIR, optical-UV (O-UV), and hard X-rays spectral bands. They are 
collected from WISE, 2MASS, GALEX, and \textit{Swift}/BAT catalogues.
The ways they were extracted from WISE and \textit{Swift}/BAT catalogues were discussed in Section \ref{sec:sample};

the analysis of 2MASS and GALEX data is presented below.

The NIR data for our objects have been obtained from the 2MASS (Two-Micron-All-Sky-Survey) extended (XSC) and point source (PSC) catalogues (\citealt{2006AJ....131.1163S}; NASA/IPAC Infrared Science Archive), the latter for objects not having matches in XSC catalogue. We used a matching radius of 10 arcsec to determine associations with our sample.

The Galaxy Evolution Explorer (GALEX) is an orbiting space telescope observing galaxies in ultraviolet light.
It imaged the sky in two ultraviolet (UV) bands, far-UV (FUV, $\lambda_{\rm eff} \sim 1528 \angstrom$), and near-UV (NUV, $\lambda_{\rm eff} \sim 2310 \angstrom$), delivering the first comprehensive sky surveys at these wavelengths.

Associations of GALEX sources with our objects have been performed using 2 arcsec matching radius. The UV fluxes were taken from the GALEX catalogue $\rm"GUVcat{\_}AIS \: GR6+7"$ \citep{2017ApJS..230...24B} as well as data from the Mikulski Archive for Space Telescopes (MAST)\footnote {https://archive.stsci.edu/missions-and-data}. We decided to use GALEX data from MAST for objects that were not reported in the $\rm "GUVcat{\_}AIS \: GR6+7"$ catalogue. We corrected for UV flux extinction in the Milky Way, using the standard relation  
\\ \\
$\rm F_{\rm \lambda} = 10^{(0.4 \times A_{\rm \lambda})} \times F_{\rm \lambda_{\rm obs}}$,
\\ \\
where $\rm A_{\rm \lambda}$ is given by
\\ \\
$\rm A_{\rm \lambda}$ = ${\Big[}\frac{A_{\rm \lambda}}{E_{\rm (B-V)}}{\Big]} \times E_{\rm (B-V)}$  = {\Bigg\{}\minibox{$8.06  E_{\rm (B-V)}$  for   $\lambda = \lambda_{\rm FUV}$\\$7.95  E_{\rm (B-V)}$ for $\lambda = \lambda_{\rm NUV}$}{\Bigg\}}
\\ \\
 and the Galactic reddenings $E_{\rm (B-V)}$ in directions to associated with our AGN UV sources are provided in the GALEX catalogue.

We found 149 NUV and 131 FUV associations for Type 1 objects and
77 NUV and 61 FUV associations for Type 2 AGN.   

Due to the limited angular resolution of GALEX, the UV radiation in Type 1 objects must be considered to be contributed not only by AGN but also by the hot stars in the host galaxies, in Type 2 objects -- only by hot stars.

For objects which did not have a valid association we assigned upper limits of their fluxes as based on detection limit of the GALEX survey in the NUV and FUV bands. These fluxes are $\rm 1.74 \times 10^{-28}\,erg\,s^{-1}$ and $\rm 3.98 \times 10^{-28}\,erg\,s^{-1}$ for the NUV and FUV bands, respectively. The number of detection is much larger than $\rm 50$ percent for Type 1 RL and RQ AGN and close to $\rm 50$ percent for Type 2 objects. Moreover, we observe that for Type 1 sources in both the NUV and FUV bands most of the undetected sources lie below the median value of all the sources. Hence the median values of our sources are statistically significant.

\begin{figure*}
\centering
\includegraphics[width=\textwidth]{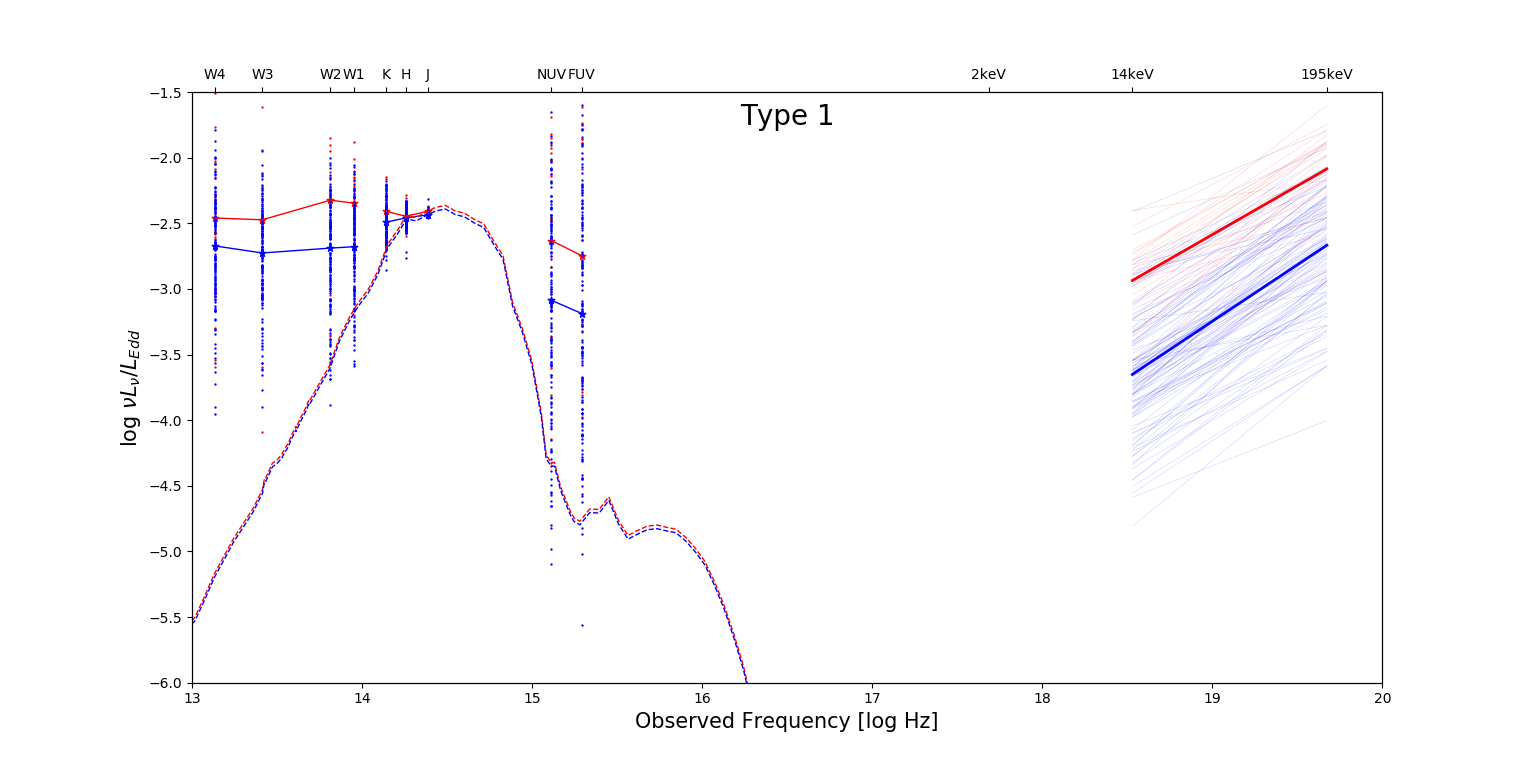}
\caption{Medians of the dominant SED components of the RL (red) and RQ (blue) Type 1 objects of the sample. All luminosities have been normalized w.r.t. to the Eddington luminosity. In case of FUV and NUV data the objects without detection have been presented with the flux limit on detection. We also present here SED template of the giant elliptical which has been scaled to K-band luminosity. Centre frequencies of each band are labelled.}
\label{img_sed_all_ty1}
\end{figure*}

\begin{figure*}
\centering
\includegraphics[width=\textwidth]{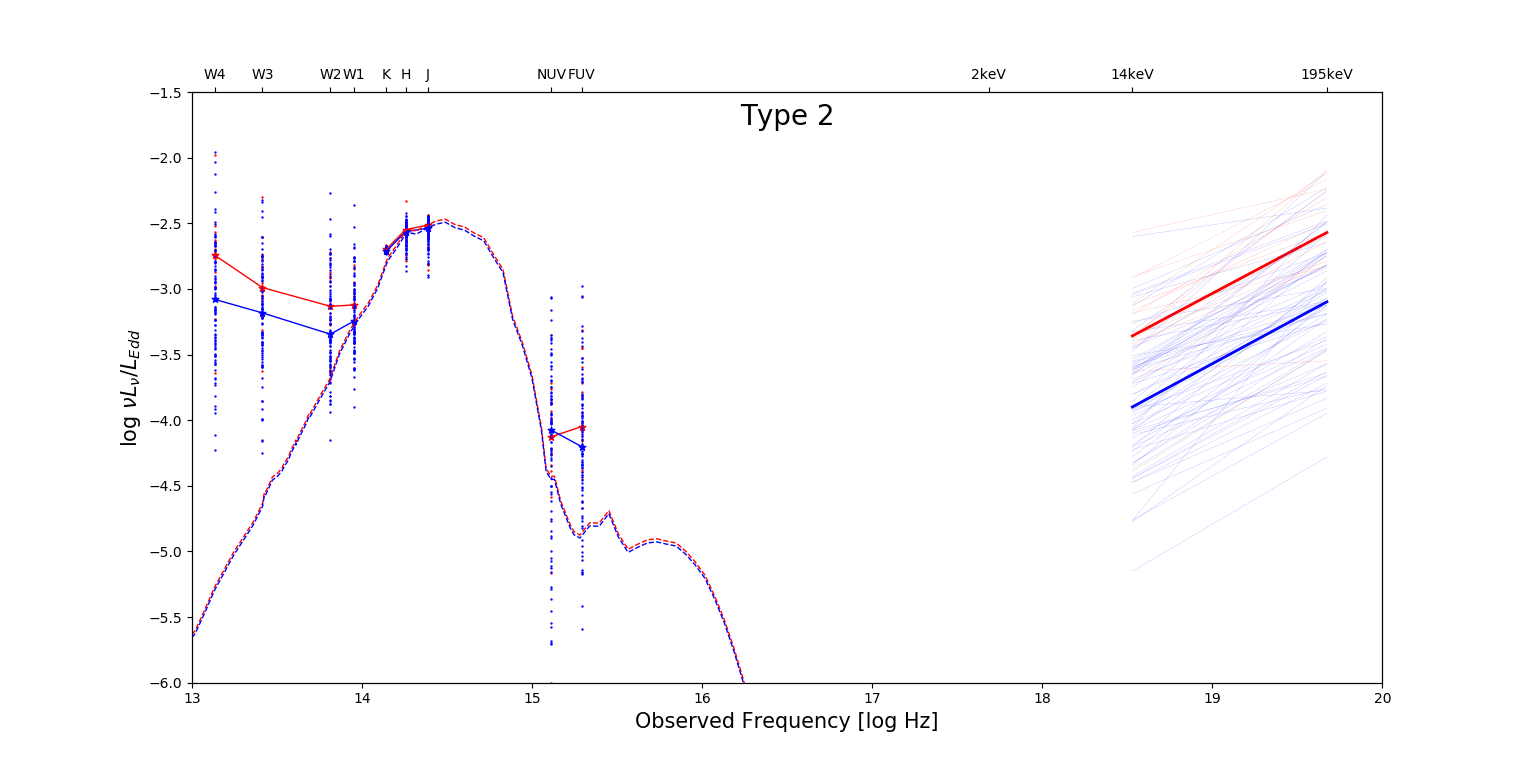}
\caption{Medians of the dominant SED components of the RL (red) and RQ (blue) Type 2 objects of the sample. All luminosities have been normalized w.r.t. to the Eddington luminosity. In case of FUV and NUV data the objects without detection have been presented with the flux limit on detection. We also present here SED template of the giant elliptical which has been scaled to K-band luminosity.  Centre frequencies of each band are labelled.}
\label{img_sed_all_ty2}
\end{figure*}

\begin{figure*}
\begin{subfigure}{0.49\linewidth}
\centering
\includegraphics[scale=0.25]{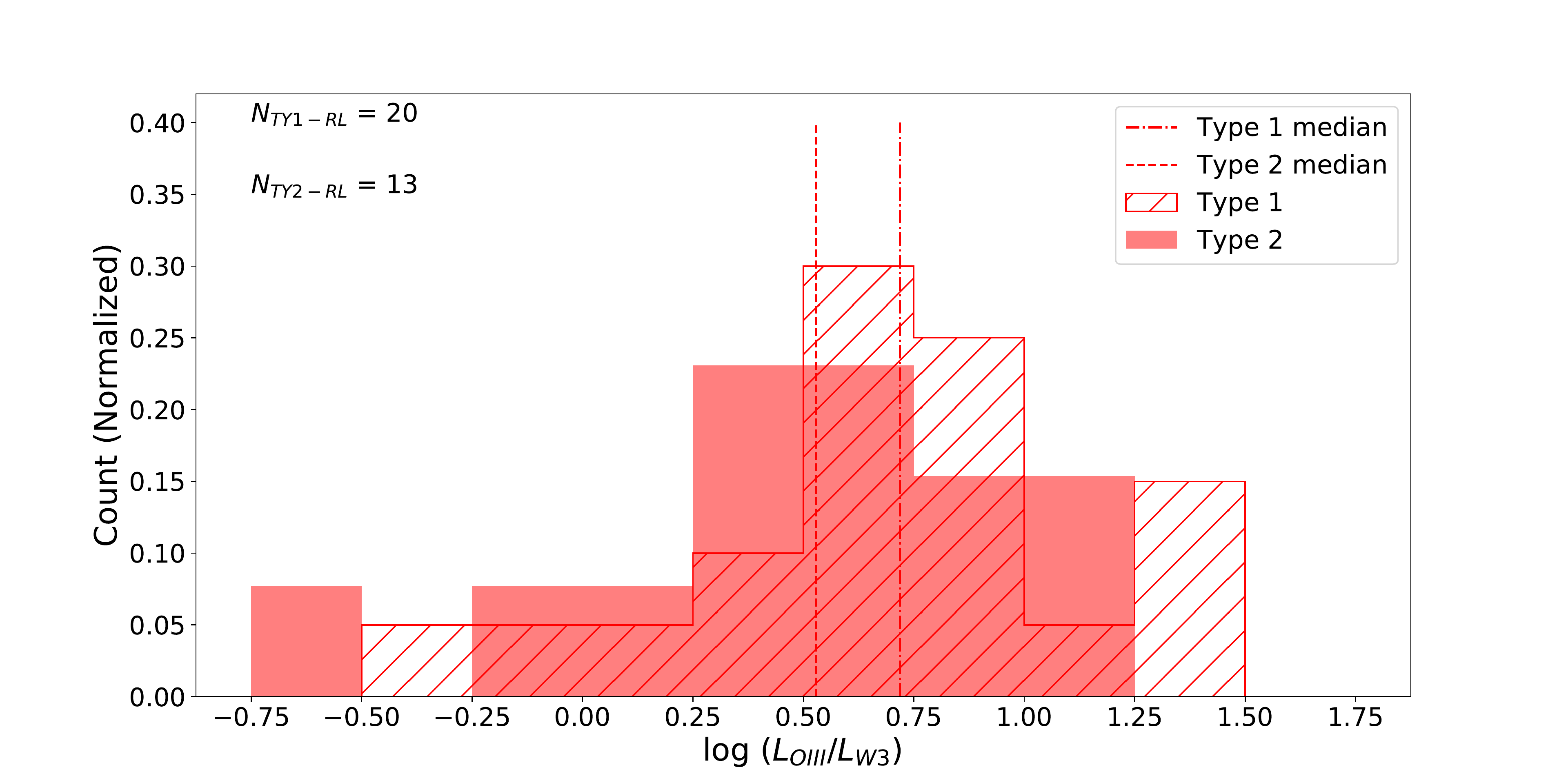}
\caption{}
\label{img_oiii_hist_rl}
\end{subfigure}\hfill
\begin{subfigure}{0.49\linewidth}
\centering
\includegraphics[scale=0.25]{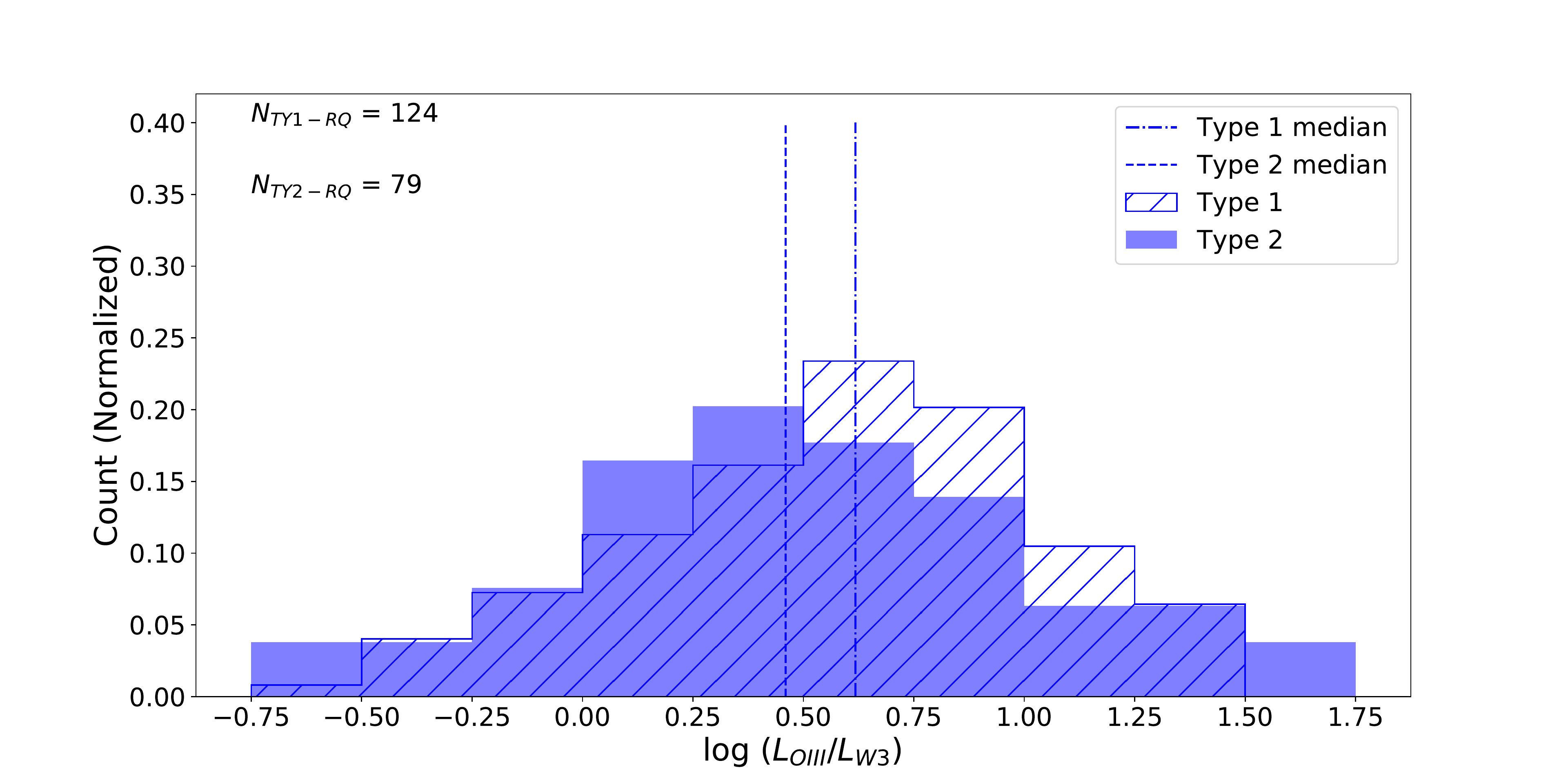}
\caption{}
\label{img_oiii_hist_rq}
\end{subfigure}\hfill]
\caption{The normalized distribution of the $\rm L_{\rm [OIII]}$ for the RL (panel a) and RQ (panel b) samples. The luminosity has been normalized w.r.t. to the W3 luminosity. We present here the distributions of the Type 1 (hatched red) and Type 2 (filled blue) subsamples of AGN. The median values of 0.72 and 0.53 for Type 1 and Type 2 RL AGN and 0.62 and 0.46 for Type 1 and Type 2 RQ AGN, respectively, are presented as dashed lines for each sample.}
\end{figure*}

The table \ref{tbl_A2} indicates total numbers of objects in our final sample that were detected per catalogue. The table consists of 290 AGN, 44 RL (27 type 1 and 17 type 2) and 246 RQ (153 type 1 and 93 type 2).

\subsection{SED / Multi-band spectra}
\label{subsec:sed_studies}

Having constructed our Type 1 and Type 2 RL and RQ subsets and determined the luminosity in different bands, we proceed to construct the composite SED for each of these samples. For each object in the sample, every available luminosity is normalized with respect to the Eddington luminosity. For objects which did not have a GALEX detection we applied an upper flux limit and determined the corresponding upper limit for luminosity.

We determine the median luminosities in MIR, NIR and hard X-ray bands. In case of the UV band, the limited number of detections causes that the medians in Type 2 AGN are close to their upper limits, in case of Type 1 AGN they are `real'.

The median hard X-ray spectra for RL and RQ AGN are constructed combining  median integrated X-ray luminosities with the median spectral slope.

We do not present in the SED figure $2-10$ keV spectra. The observed spectra in this band are curved  and presenting all of them together  in the visible fasion is rather impossible. Intrinsically they are expected to provide extension of BAT spectra with similar slopes as the BAT ones to lower energies.

We plot also in these figures templates of the giant ellipticals. They are
scaled to be equal in the K band to the median luminosities of the RL and RQ samples.

The composite SEDs of  Type 1 and Type 2 AGN are presented in Fig. \ref{img_sed_all_ty1} and Fig. \ref{img_sed_all_ty2}, respectively. All luminosities are calculated ignoring K-corrections and, hence, presented as a function of the observed frequency. As we can see  the SEDs of our objects are contributed not only by the AGN, but also by stars of galaxies hosting those AGN. The latter dominate the radiation in the NIR. And in the case of type 2 AGN and very low
Eddington ratio Type 1 AGN they dominate in the optical bands as well. The non-AGN contributions can arise also in other SED portions, in particular in the MIR and UV band as provided there by star formation regions (SFRs). However their significance in the samples studied by us is rather minor: in the MIR band -- because SFRs are predicted to produce much redder MIR spectra than observed in the objects of our sample \citep[e.g.][]{2017ApJ...835...74I,2018ApJ...866...92L}; and in the UV band only in Type 2 AGN -- otherwise domination
of UV radiation by SFR should be similar in Type 1 and
Type 2 AGN, whereas our Type 1 objects are more UV luminous than 
Type 2 objects (see Fig. \ref{img_sed_all_ty1} and \ref{img_sed_all_ty2}).
Hence the SEDs of Type 1 AGN in our samples are dominated by three components, 
the MIR, UV and hard X-ray ones. Their relative contributions
to the total observed luminosities are of the same order.
This concerns both RL and RQ AGN, and statistically
significant differences appear only in the X-ray band, where the RL AGN
are $2$ times more luminous than the RQ ones.

\begin{table}
\begin{center}
\caption{Number of objects in our final sample that were detected per catalogue.}
\label{tbl_A2}

\begin{tabular}{ c c c c c }
\hline
  Catalogue & \multicolumn{2}{c}{Type 1} & \multicolumn{2}{c}{Type 2} \\
  \hline\hline
 - & RL & RQ & RL & RQ \\
 \hline
Total & 27 & 153 & 17    & 93 \\
\hline
WISE & 27 & 153 & 17 & 93 \\
2MASS & 27 & 153 & 17 & 93 \\
Hard X-ray & 27 & 153 & 17 & 93 \\
GALEX FUV & 20 & 111 & 12 & 49 \\
GALEX NUV & 22 & 127 & 13 & 64 \\
OIII & 20 & 124 & 13 & 79 \\
\hline
\end{tabular}
\end{center}
\end{table}

\section{DISCUSSION}
\label{sec:discussion}

The hard X-ray emitting AGN with black hole masses $\rm > 10^{8.5}\,M_{\odot}$ selected by us are found
to accrete at the rates corresponding to Eddington ratio -- if not accounting for a few outliers -- enclosed between $0.001$ and $0.03$ (see Appendix \ref{app:bol} and Fig. \ref{img_hist_lam_e}). They have 
bimodal distribution of radio loudness and a fraction of radio-loud objects 
is about $0.15$. Further analysis of our sample including the radio morphologies of the sources, their host galaxies and their environments will be explored in our future works.
The observed radiative output of RL and RQ AGN is dominated by 
MIR and X-ray components and the only significant difference between them is 
that the RL AGN are on average $\sim$two times X-ray louder than their RQ counterparts. Possible interpretations of these results are discussed below.

\subsection {The origin of hard X-rays in RQ and RL AGN}
\label{subsec:origin}
Production of hard X-rays is commonly interpreted as a result of comptonization
of optical-UV radiation of cold accretion disk by hot electrons 
($\rm kT_{\rm e} \sim 100$\,keV) in optically thin coronas \citep[][and refs. therein]{2018MNRAS.480.1247K}. Such coronas are likely heated by magnetic reconnection 
\citep{1979ApJ...229..318G,2017ApJ...850..141B}. Another possibility is that hot 
coronas represent central portions of accretion flows which are predicted by 
some models to result from conversion of the cold, optically  thick and 
geometrically thin disks into hot, optically thin and geometrically thick 
inflows \citep{2002A&A...392L...5M,2018ApJ...854....6H}. Such models
are often applied  for AGN in which  X-ray reflection features are weak
and the radiative output is not dominated by UV bump
\citep[e.g.][]{2010MNRAS.402..724P,2019MNRAS.484..196T,2019ApJ...870...73Y}.  
However noting that weaker reflection features may result also from
higher ionization of accretion disks in such AGN, while their relatively
lower UV luminosities can come from extinction associated with the polar dust
(see section \ref{subsec:polar}), truncation of cold  accretion disks may not be required. 

Interpretation of the origin of hard X-ray  is even more uncertain in case 
of RL AGN. Their larger X-ray loudness than of RQ ones may suggest 
that X-rays in these objects are partially contributed by the 
jet base \citep[e.g.][]{1998MNRAS.299..449W}. However, as we noticed in our previous 
paper \citep{2018MNRAS.480.2861G}, despite the difference in the X-ray loudness,
the spectral slopes and high energy breaks are very similar in 
RL and RQ AGN selected to have similar BH masses and Eddington 
ratios. We used this as an argument in favor of the same location and same
emission mechanism of hard X-ray sources in RL and RQ AGN, the location being 
the hot corona in the innermost portions of accretion flow and the mechanism --
Compton up-scattering of lower energy disk photons. In this paper we explore 
such interpretation by  taking advantage of having in our samples Type 1 and 
Type 2 AGN which  allows us to verify dependence  of the X-ray radiation 
luminosity on the  inclination angle. We carried out such verification
by comparing distributions of X-ray--to--MIR luminosity (Fig. \ref{img_hist_hxlum}, Fig. \ref{img_x_hard_hist_rl} and Fig. \ref{img_x_hard_hist_rq}). They are found to be
very similar for Type 1 and type 2 AGN for both RL and RQ AGN, which
suggests that the departure of  hard X-rays from the isotropy is statistically 
insignificant. The isotropy indicates also that X-ray sources in 
both RQ and RL AGN cannot be too compact, otherwise their isotropy would be 
strongly affected by gravitational lensing. If trying to 
interpret the larger X-ray loudness of RL AGN by the jet contribution,
this would require tight tuning of 
parameters discribing  the geometry and kinematics of a jet. Furthermore, 
production of X-rays by a slow  jet with similar luminosity as by accretion
flow would imply that most of the jet energy required to efficiently power 
the radio lobes is dissipated and radiatively lost already at the base.  
Hence, we are tempted to speculate that in both RQ and RL AGN production 
of X-rays is dominated by hot coronae associated with central portions
of accretion flows and that 
more efficient X-ray  production in RL AGN result from larger 
magnetization of innermost portions of accretion flows and from larger BH spins
which are required to afford efficient jet production in the MAD scenario
involving the Blandford-Znajek mechanism.

\subsection {UV luminosities and the `polar' dust}
\label{subsec:polar}
Having reprocessed  UV radiation of accretion disk to MIR radiation
in dusty  molecular tori  and assuming absence of dust in ionization cones
(the zone not protected against the UV radiation by the torus) 
one might expect to get the ratio of MIR to O-UV luminosity of the order CF $\rm \simeq
N_{\rm Type 2}/(N_{\rm Type 1}+N_{\rm Type 2})$, where $\rm N_{\rm Type 1}$ and $\rm N_{\rm Type 2}$ are 
the numbers of Type 1 AGN and Type 2 AGN, respectively \citep[][and refs therein]{2016MNRAS.461.2346G}. We note that in our sample
the latter ratio is 0.38, while integrated mid-IR luminosity reaches or even 
exceeds  UV luminosities of Type 1 AGN. This strongly implies that a significant 
fraction of UV radiation is extincted and  reprocessed into IR radiation by 
the dust located within the ionization cone. Presence of such, commonly called, 
polar dust is theoretically predicted to be common in the  AGN accreting at 
moderate rates, and  this is because at such rates the pressure of UV 
radiation is too small to protect the ionization zone against the dust
\citep{2013ApJ...771...87H,2017ApJ...838L..20H,2017Natur.549..488R,2018ApJ...866...92L}. Existence of the polar dust is observationally confirmed 
by mid-IR interferometric observations \citep{2009A&A...495..137H,2009A&A...502..457G,2011A&A...536A..78K,2011A&A...531A..99T,2016A&A...591A..47L}.
They show the extension of mid-IR images along the ionization cones.  

Comparing the UV fluxes of our RL and RQ Type 1 AGN (Fig. \ref{img_sed_all_ty1}) we can see that UV is less extincted in RL objects. This may result from smaller amount of dust in the vicinity of powerful jets. However this result is statistically weak and need confirmation by future studies using larger samples. 

We also compared luminosities of [OIII] lines. We make use of the [OIII] emission line flux from \cite{2017ApJ...850...74K} derived from the SDSS spectra. For both RL and RQ AGN they are found on average stronger in Type 2 objects than in Type 1 AGN (Fig. \ref{img_oiii_hist_rl} and \ref{img_oiii_hist_rq}). This difference can be explained by the polar gradient of the dust distribution within the ionization zone. Since the fraction of detection of [OIII] is larger than 50 percent for both Type 1 and Type 2 RL and RQ AGN, the median values of our sources are real for the entire sample (see Section \ref{sec:sed_distribution}).

\section{SUMMARY}
\label{sec:summary}

We summarize our results as follows:

-- AGN selected from the \textit{Swift}/BAT catalogue with black hole masses 
   $\rm >10^{8.5}\,M_{\odot}$ have bimodal distribution of radio loudness.
   Medians of radio loudness distribution  of RL and RQ AGN differ 
   by a factor $429$, and  the radio-loud-fraction is $\sim 0.15$; 

-- The only statistically signifcant  difference between SEDs of RL and RQ
   AGN are larger X-ray luminosities in the former class.  
   The isotropy of X-ray luminosities implied by our study in both RL and RQ AGN 
   from finding that hard X-ray-to-MIR luminosity ratios in Type 1 and Type 2 AGN
   do not differ seems to disfavor explanation of this difference by 
   postulating the jet contribution to hard X-rays in RL AGN. 
   Together with previous findings that slopes and  high energy breaks of their hard X-ray spectra 
   are on average very similar in both RL and RQ AGN, it appears to support our 
   premise that in both samples the hard X-ray are produced in hot central portions of 
   accretion flows and that luminosity difference can be associated with
   having faster rotating BHs and larger magnetic fluxes in AGN producing
   jets; 
       
-- The radiative output of both RL and RQ AGN is dominated by radiation
   in the MIR and hard X-ray bands.  Their observed UV luminosities are  
   likely to be suppressed due to extinction. Lower  UV luminosities 
   as compared with those predicted by standard accretion disk models can 
   also result from the accretion disk truncation. However this alone
   cannot explain MIR luminosities larger than $\rm CF \times  L_{\rm UV,em}$.

\section*{Acknowledgements}

We thank Greg Madejski for his comments which helped us improve the paper. We acknowledge financial support by the Polish National Science Centre grants 2016/21/B/ST9/01620 and 2017/25/N/ST9/01953.







\appendix

\section{$M_{\rm BH}$ Estimations}
\label{app:mbh}

As it is stated in Section \ref{sec:intro}, BH masses of all AGN in our sample are calculated
using method which is based on the relation between luminosities of host 
galaxies and black hole masses \citep{2003ApJ...589L..21M,2007MNRAS.379..711G}.
We adopted the formula from \cite{2007MNRAS.379..711G}
\\

$\rm \log(M_{\rm BH}/M_{\odot}) = -0.37(\pm 0.04)(M_{\rm K} + 24) + 8.29(\pm 0.08),$
\\

\noindent where $\rm M_{\rm K}$ is the absolute K-band magnitude of the galaxy.

\begin{figure}
\centering
\includegraphics[scale=0.25]{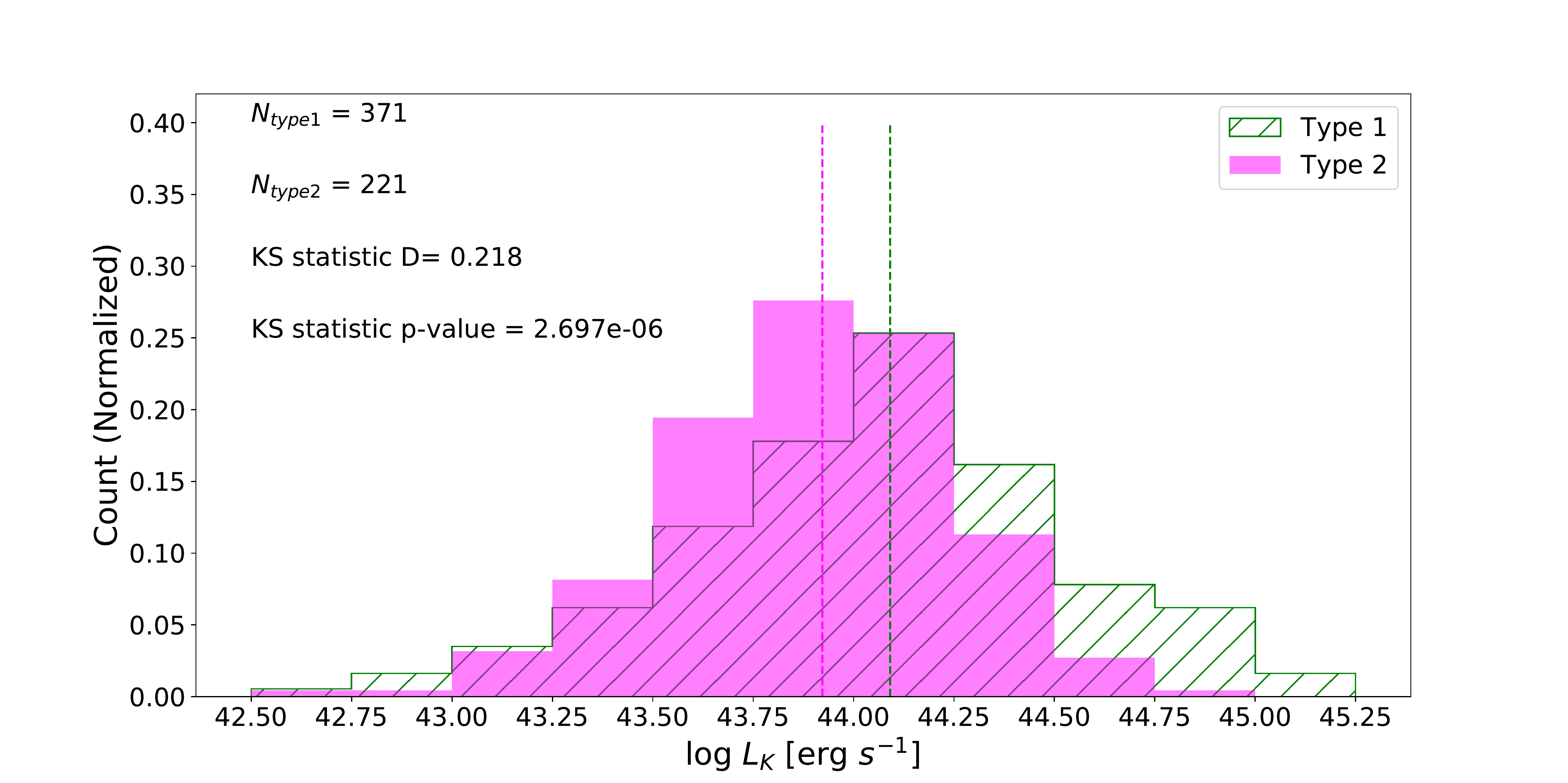}
\caption{The histogram of the observed K-band luminosity of the Type 1 (hatched green) and Type 2 (filled pink) samples. Also shown are the median values of both samples: 44.09 for Type 1 and 43.92 for Type 2 AGN.}
\label{img_hist_klum_obs}
\end{figure}

\begin{figure}
\centering
\includegraphics[scale=0.25]{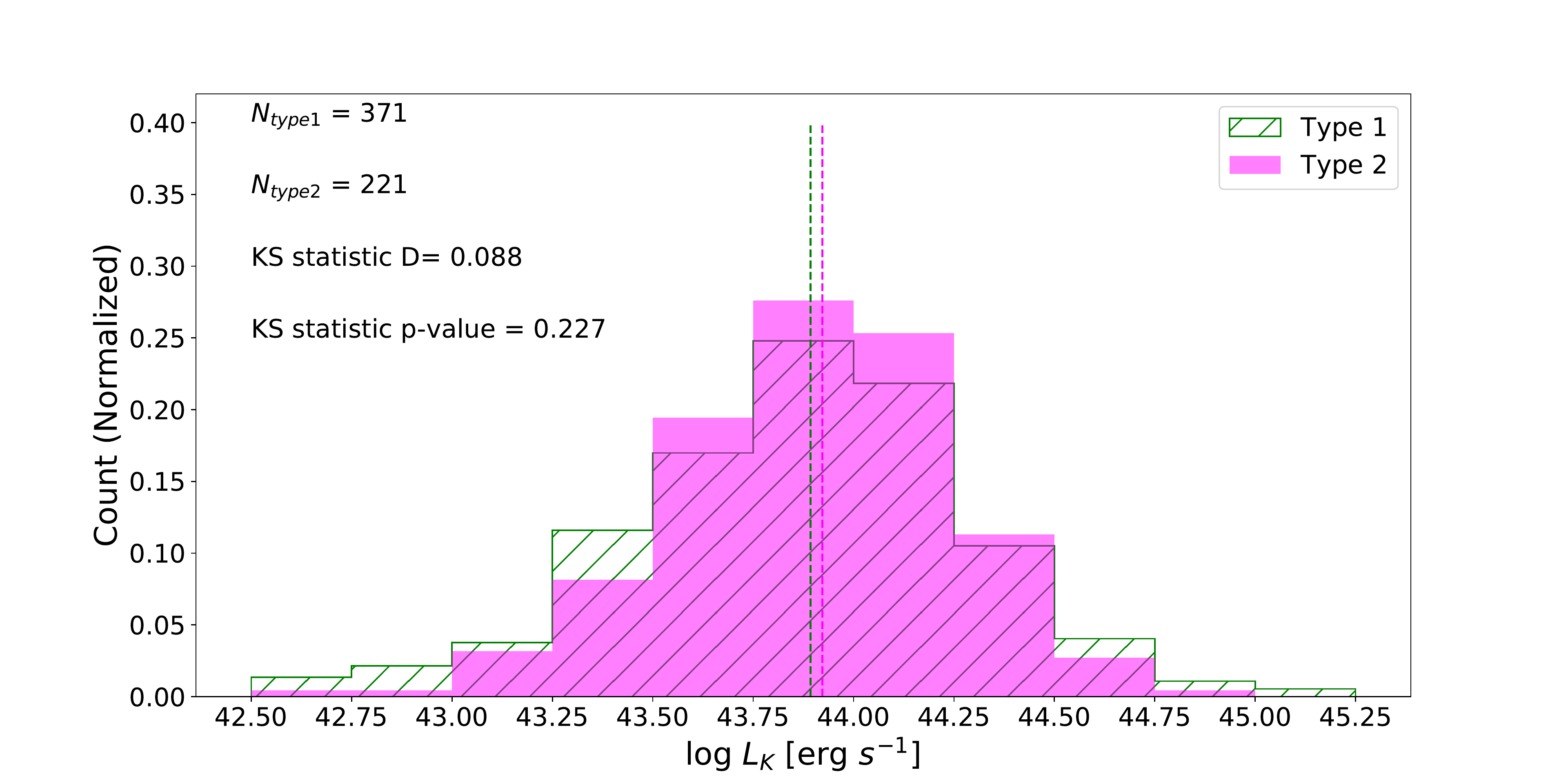}
\caption{The histogram of K-band luminosity for Type 1 objects the luminosity is derived from the galaxy template and the observed luminosity in J-band, for Type 2 objects from the observed luminosity. The median values are 43.89 for Type 1 AGN (hatched green) and 43.92 for Type 2 AGN (filled pink).}
\label{img_hist_klum_gal}
\end{figure}

\begin{figure}
\centering
\includegraphics[scale=0.25]{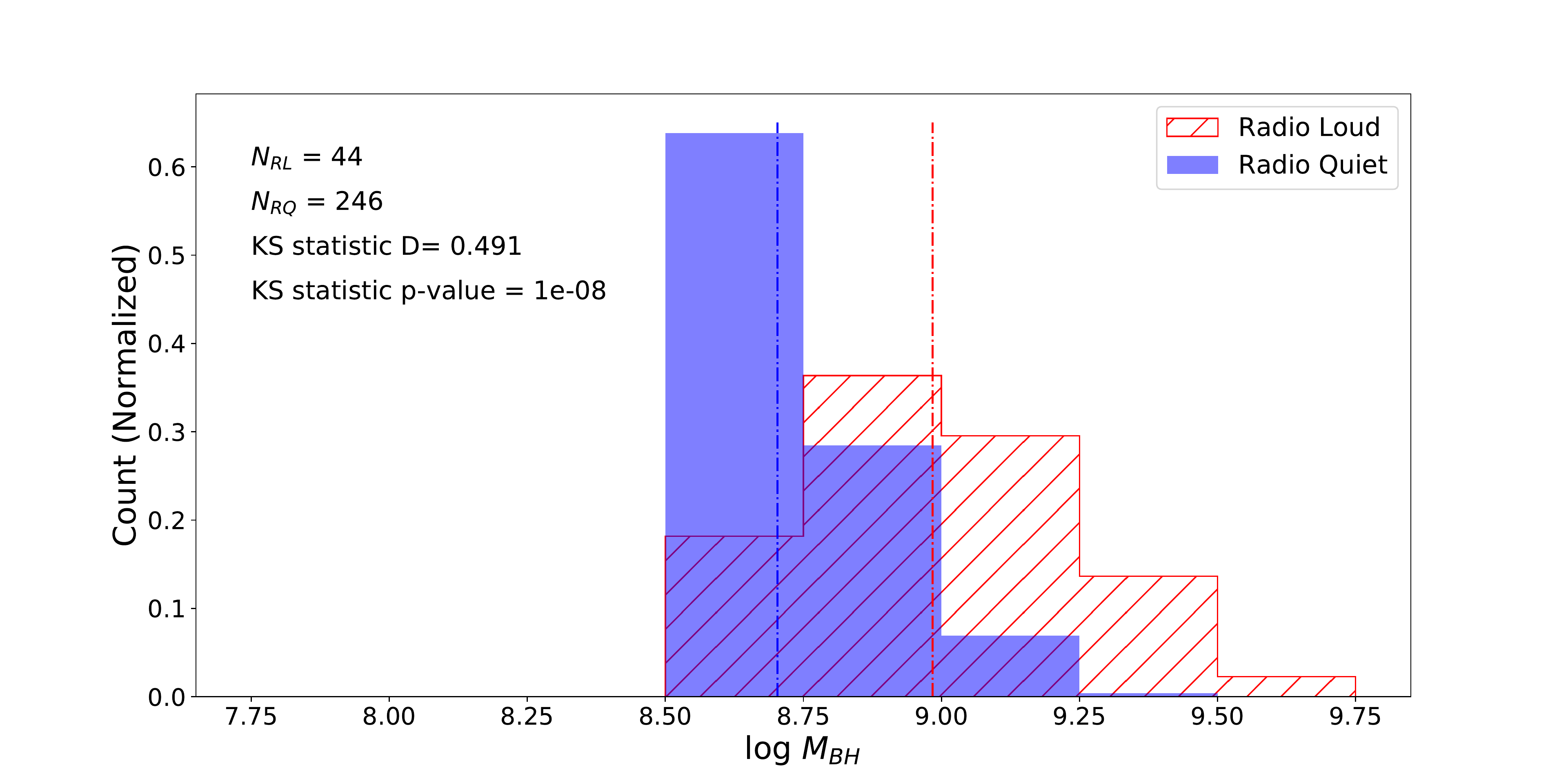}
\caption{The normalized distribution of the black hole masses of RL (hatched red) and RQ (filled blue) AGN with median values of 8.9 and 8.7, respectively, presented as dashed lines.}
\label{img_mbh_dist}
\end{figure}

This method can be directly applied for Type 2 AGN, where AGN contribution
to near-IR is negligible if any. Using this formula for Type 1 AGN needs to be assured that contribution of AGN radiation to near-IR is much smaller than of the host galaxy. While it is not the case in high accretion rate objects such as quasars, for AGN accreting at rates corresponding with Eddington ratio smaller than $0.03$ the domination of the near-IR luminosities 
by the host galaxies is well documented \citep{2010MNRAS.402..724P,2012MNRAS.423..600S}. Nevertheless since some contribution from  AGN is unavoidable, in order to minimize overestimation of BH masses in Type 1 AGN by ignoring it, the K-luminosity used in Graham's formula is not taken directly from observations but derived from J-luminosity using the template of giant elliptical \citep{1998ApJ...509..103S}.

This leads to lowering 
the BH mass overestimation because AGN contribution  to near-IR  
is larger in K-band than in J-band. 
Following this procedure the difference between median values of K-luminosity 
distributions in Type 1 and Type 2 AGN dissapears (see Fig. \ref{img_hist_klum_obs} and \ref{img_hist_klum_gal}). This is as predicted by the Unified Scheme \citep{1995PASP..107..803U} and having strongly dominated near-IR
luminosity by host galaxies.

Fig. \ref{img_mbh_dist} shows the black hole mass distribution of RL and RQ AGN. Also presented in the figure are the median values of each population.

\section{Bolometric luminosities and Eddington ratio}
\label{app:bol}

The SEDs of Type 1 AGN are usually dominated by optical-UV, hard X-ray, and
MIR components, where the latter can be contributed by dust in the molecular 
torus as well, as by the polar dust. When amount of the latter is too small
to extinct significantly optical-UV radiation, then the bolometric luminosity, 
defined to be equal to amount of energy produced 
by the accretion flow per unit of  time,  is a sum of only two first components.
Situation complicates if extinction by the polar dust is not negligible and
leads to conversion of some portion of optical-UV radiation to the IR-band.
In such a case the bolometric luminosity is given by formula
$\rm L_{\rm bol} \simeq L_{\rm {O-UV,disk}} + L_{\rm X,corona}$
, where $\rm L_{\rm O-UV,disk} = L_{\rm O-UV} +  L_{\rm IR,polar}$, and $\rm L_{\rm IR,polar}$ is the infrared 
radiation produced in the polar zone. Since the observed MIR luminosity
is the sum of MIR emitted  in the polar region and in the torus,
in order to calculate $\rm L_{\rm bol}$ we need to know what fraction of the total 
IR luminosity is produced in the polar zone. This could be evaluated from
extinction measurements  using decrements of narrow emission lines. However
the fraction of objects in our sample for which such data are available 
is very small. Fortunately, we can avoid this problem if noting that the 
fraction of optical-UV radiation converted to MIR in the circumnuclear 
molecular torus is $\rm CF \simeq N_{\rm Type 2}/(N_{\rm Type 1}+N_{\rm Type 2})$.
Then noting that
infrared luminosity produced by torus is $\rm L_{IR,torus} = CF \times L_{\rm O-UV,disk}$ and that
$\rm L_{\rm IR,polar} = L_{\rm O-UV,disk} - L_{\rm O-UV}$ the bolometric luminosity can be 
approximated by formula
$$\rm L_{\rm bol} = L_{\rm X} + \frac{(L_{\rm O-UV} + L_{\rm IR})}{(1+CF)} \, .$$
Then the bolometric luminosities of all objects in our sample of Type 1 AGN
can be calculated individually using following approximations

$\rm L_{\rm X} = L_{14-195} \frac{1-(1/195)^{2-\Gamma_{ph}}}{ 1-(14/195)^{2-\Gamma_{ph}}} $,
\\ 

\noindent (Note that the intrinsic hard X-ray spectra are approximated by extension of the 14-195 keV BAT spectra down to 1 keV),
\\

$\rm L_{\rm IR} = 3.1 \times \nu_{w3} L_{\nu_{w3}}$,
\\ 

$\rm L_{\rm O-UV} = 3.1 \times \nu_{FUV} L_{\nu_{FUV}}$. 
\\ 

Note that same values of numerical
coefficients in formula for $L_{\rm IR}$ and for $L_{\rm O-UV}$ are
accidental. For the IR luminosity it comes from assumption that the IR spectrum
have an energy flux index $\rm \alpha_{IR}=1$ $\rm (F_\nu \propto \nu^{-\alpha_{IR}})$ and is
enclosed between $\rm 1 \upmu m$ and $\rm 22 \upmu m$, for the O-UV luminosity adopting the
template of the spectrum in this band from \cite{2004MNRAS.351..169M}.

\begin{figure}
\centering
\includegraphics[scale=0.25]{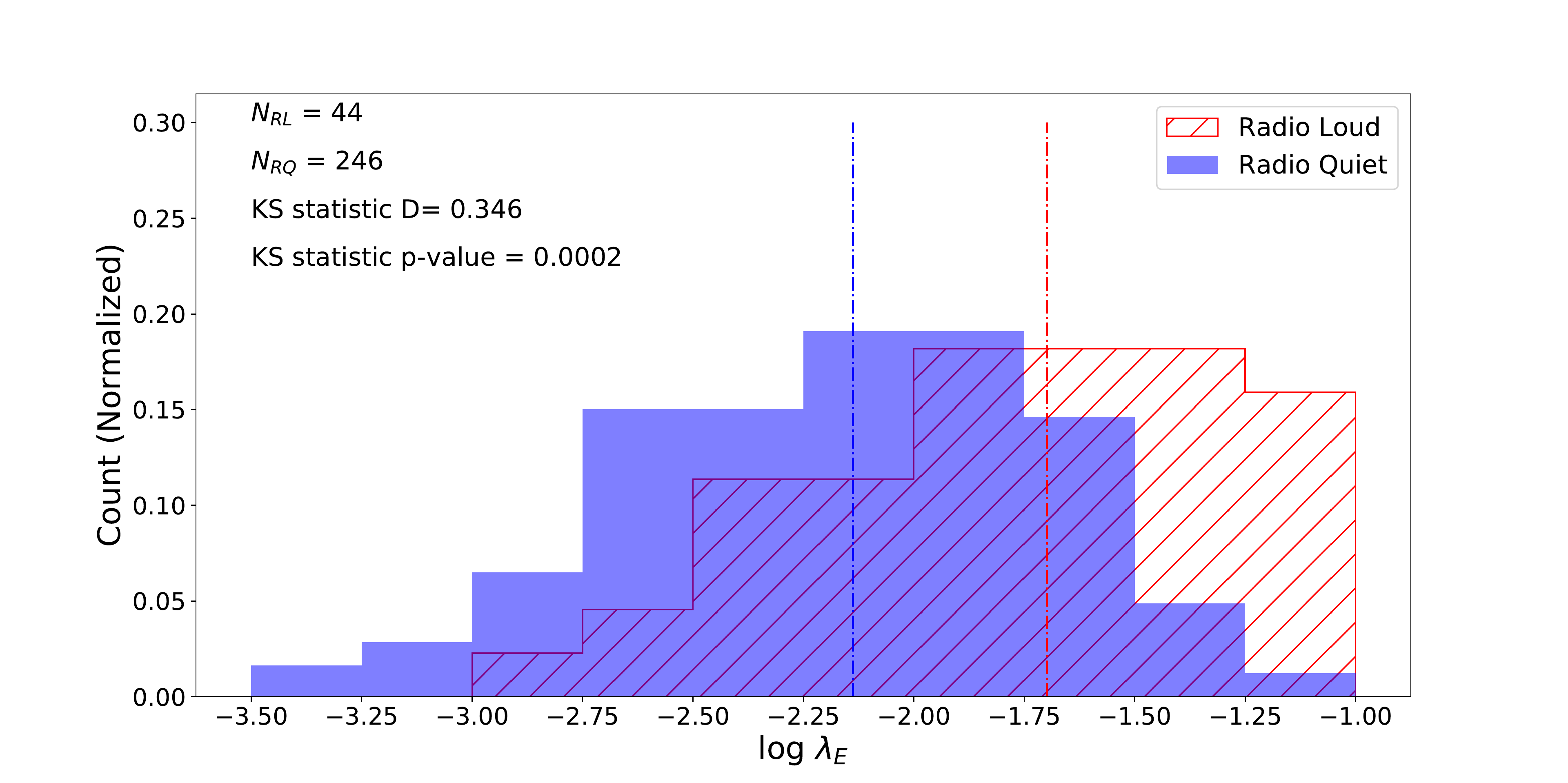}
\caption{The histogram of Eddington ratio, where the RL sample is represented by hatched red and RQ sample by filled blue with median values of $-1.7$ and $-2.14$, respectively, presented as dashed lines.}
\label{img_hist_lam_e}
\end{figure}

The distribution of the Eddington ratio, $\lambda_{\rm E} = L_{\rm bol}/L_{\rm Edd}$, for the whole sample is shown on Fig. \ref{img_hist_lam_e} and is within the range of $0.001 \leq \lambda_{\rm E} \leq 0.03$. Also presented in the figure are the median values of each population.

\section{Redshift selected sample}
\label{app:red}

\begin{figure}
\centering
\includegraphics[scale=0.25]{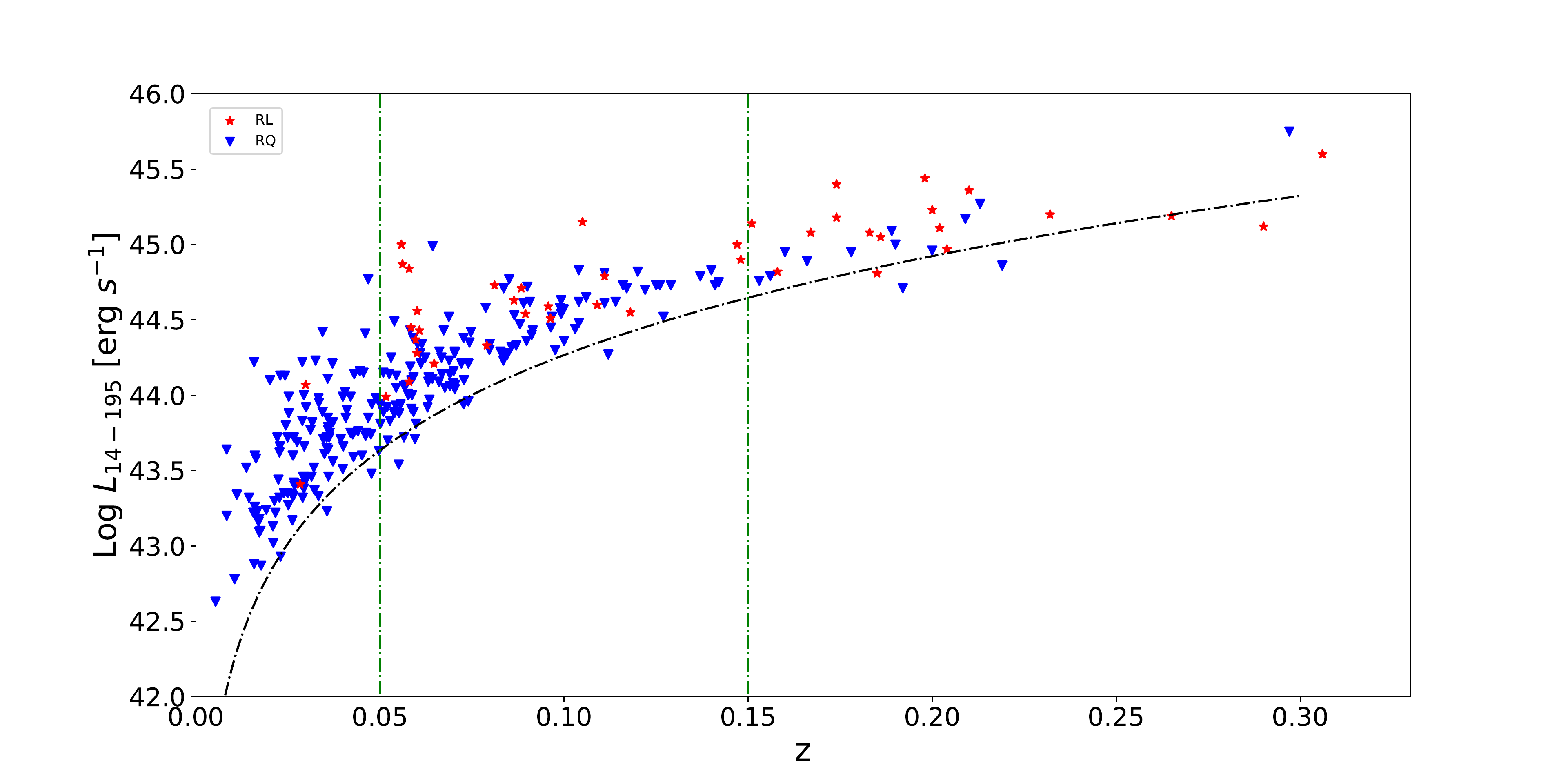}
\caption{The distribution of hard X-ray luminosity versus redshift of RL (red stars) and RQ (blue triangles) subsamples of AGN. The redshift range of $0.05\leq z\leq0.15$ which will be used to construct a redshift limited subsample is marked. }
\label{img_z_vs_lumX}
\end{figure}

\begin{figure}
\centering
\includegraphics[scale=0.25]{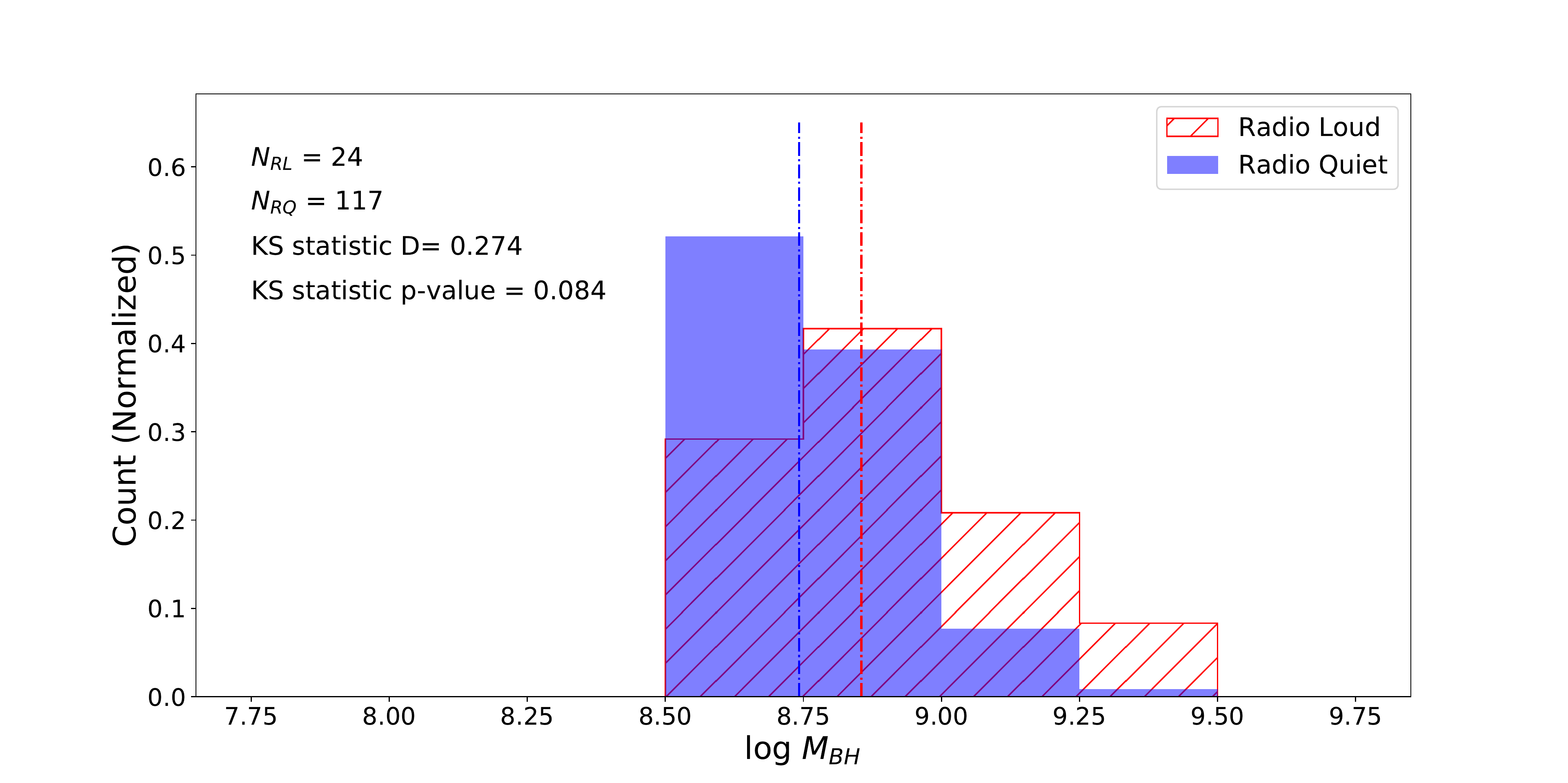}
\caption{The normalized distribution of the black hole masses of a redshift limited sample of RL (hatched red) and RQ (filled blue) AGN with median values of 8.8 and 8.7, respectively, presented as dashed lines. Compare with Fig. \ref{img_mbh_dist}.}
\label{img_mbh_dist_z_cut}
\end{figure}

\begin{figure}
\centering
\includegraphics[scale=0.25]{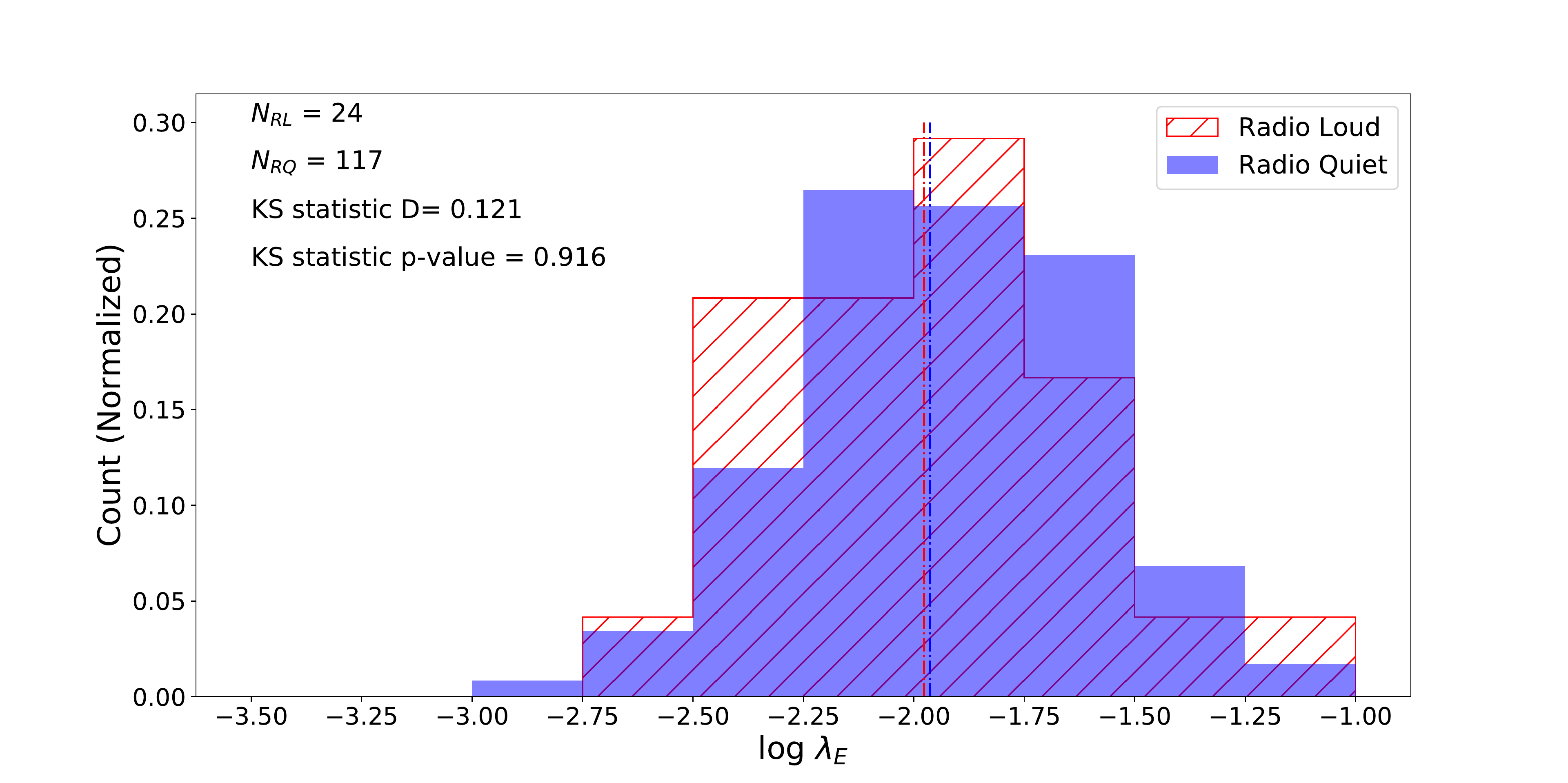}
\caption{The normalized distribution of the Eddington ratio for the redshift limited sample of RL (hatched red) and RQ (filled blue) AGN with median values of $-1.98$ and $-1.996$, respectively, presented as dashed lines. Compare with Fig. \ref{img_hist_lam_e}.}
\label{img_hist_lam_e_z_cut}
\end{figure}

\begin{figure}
\centering
\includegraphics[scale=0.25]{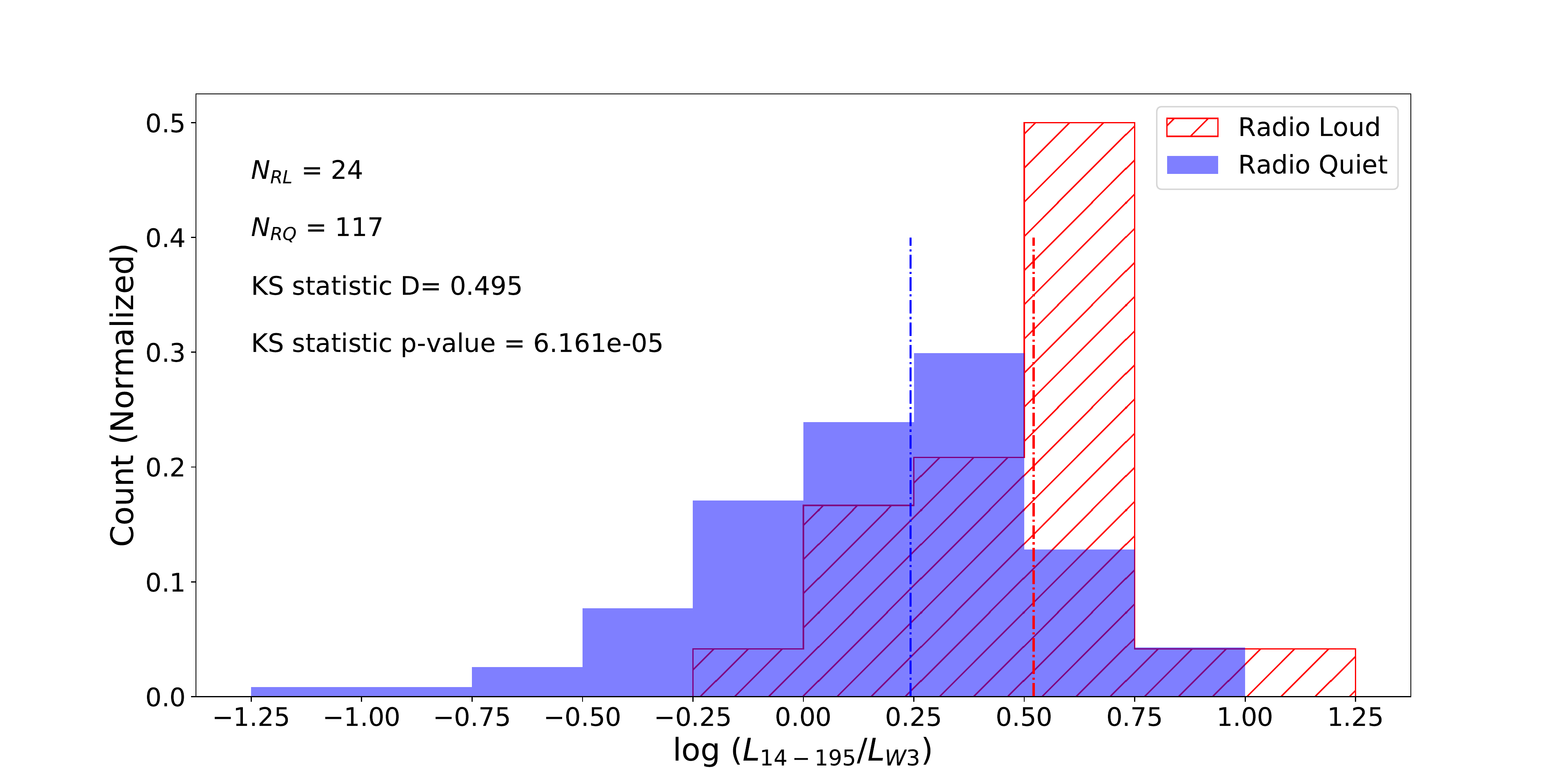}
\caption{The normalized distribution of the X-ray loudness of a redshift limited sample of RL (hatched red) and RQ (filled blue) AGN with median values of 0.52 and 0.24, respectively, presented as dashed lines. Compare with Fig. \ref{img_hist_hxlum}.}
\label{img_hist_hxlum_z_cut}
\end{figure}

Results of our comparisons of RL and RQ AGN are expected to be biased by
having them selected from pretty broad range of redshift ($z \leq 0.35$).
Since RQ AGN have larger extension of Eddington ratios and of BH masses
to lower values than RL AGN, some differentiation of radiative properties
of RL and RQ objects can be contributed by having on average RL AGN more
luminous than RQ AGN. We verify whether this can explain the difference
between X-ray loudness of RL and RQ AGN by limiting our samples to the
redshift range $0.05 - 0.15$ (Fig. \ref{img_z_vs_lumX}). As we can see comparing Fig. \ref{img_mbh_dist_z_cut} and \ref{img_hist_lam_e_z_cut}
with Fig. \ref{img_mbh_dist}  and \ref{img_hist_lam_e}, within this redshift range RL and RQ AGN have very
similar distributions and ranges of BH masses and Eddington ratios. For these samples the difference in X-ray
loudness dropped, but only by a factor 1.16 (from 2.2 down to 1.9, compare Fig. \ref{img_hist_hxlum} and \ref{img_hist_hxlum_z_cut}). This indicates that the difference in X-ray loudness between RL and RQ AGN is real rather than imposed by the selection effects.

\section{Catalogue}

The complete catalogue is available as supplementary material online. Table \ref{tbl_A1} gives a brief description of the columns in the catalogue. \\

\begin{table*}
\begin{center}
\caption{Description of the RL \& RQ catalogue. The catalogue is available as supplementary material online.}
\label{tbl_A1}

 \begin{tabular}{c c c} 
 \hline
 Column Name & Unit  & Description \\ 
 \hline\hline
Name &  & Counterpart name from \textit{Swift}/BAT catalogue of AGN \\
RA & deg & Counterpart RA coordinates from \textit{Swift}/BAT catalogue of AGN\\
DEC & deg &   Counterpart DEC coordinates from \textit{Swift}/BAT catalogue of AGN \\
z & & Redshift\\
$\rm \log\_R$ & & Log of the radio loudness\\
Radio\_class & & Radio classification of object (RL/RQ)\\
Optical\_type & & Optical classification\\
log\_MBH & $\rm M_{\odot}$ & BH mass determined from K-band luminosity\\
log\_Eddington\_ratio & & Eddington ratio\\
log\_NH & $\rm cm^{-2}$ & Column density\\
$\rm \log\_L\_W1$ & $\rm erg\,s^{-1}$ & WISE Luminosity at $\rm \nu_{\rm W1}$\\
$\rm \log\_L\_W2$ & $\rm erg\,s^{-1}$ & WISE Luminosity at $\rm \nu_{\rm W2}$\\
$\rm \log\_L\_W3$ & $\rm erg\,s^{-1}$ & WISE Luminosity at $\rm \nu_{\rm W3}$\\
$\rm \log\_L\_W4$ & $\rm erg\,s^{-1}$ & WISE Luminosity at $\rm \nu_{\rm W4}$\\

$\rm \log\_L\_J$ & $\rm erg\,s^{-1}$ & 2MASS Luminosity at $\rm \nu_{\rm J}$\\
$\rm \log\_L\_H$ & $\rm erg\,s^{-1}$ & 2MASS Luminosity at $\rm \nu_{\rm H}$\\
$\rm \log\_L\_K$ & $\rm erg\,s^{-1}$ & 2MASS Luminosity at $\rm \nu_{\rm K}$\\

$\rm \log\_L\_FUV$ & $\rm erg\,s^{-1}$ & GALEX Luminosity at $\rm \nu_{\rm FUV}$\\
$\rm Upper\_limit\_FUV$ &  & $\rm \log\_L\_FUV$ presented is an upper limit\\
$\rm \log\_L\_NUV$ & $\rm erg\,s^{-1}$ & GALEX Luminosity at $\rm \nu_{\rm NUV}$\\
$\rm Upper\_limit\_NUV$ &  & $\rm \log\_L\_NUV$ presented is an upper limit\\

$\rm \log\_L\_2\_10$ & $\rm erg\,s^{-1}$ & X-ray Luminosity in the $2-10$ keV band\\
$\rm \log\_L\_14\_195$ & $\rm erg\,s^{-1}$ & X-ray Luminosity in the $14-195$ keV band\\
Gamma\_14\_195 &  &  Photon index in the X-ray band $14-195$ keV\\

 \hline
\end{tabular}
\end{center}
\end{table*}

\bsp	
\label{lastpage}
\end{document}